\documentclass[prx,twocolumn,amsmath,amssymb,superscriptaddress,eqsecnum]{revtex4-2}
\usepackage{graphicx,amsmath,relsize,epstopdf,upgreek,color,mathtools,bm,times,rotating}
\usepackage[T1]{fontenc}
\usepackage[hyphenbreaks]{breakurl}
\usepackage[colorlinks=true,linkcolor=blue,citecolor=blue,urlcolor =blue]{hyperref}

\newcommand{\ket}[1]{|{#1}\rangle}
\newcommand{\bra}[1]{\langle{#1}|}
\newcommand{\inner}[2]{\langle{#1}|{#2}\rangle}
\newcommand{\rvec}[1]{\pmb{#1}}
\newcommand{\dyadic}[1]{\pmb{#1}}
\newcommand{\tr}[1]{\mathrm{tr}\!\left\{#1\right\}}
\newcommand{\D}{\mathrm{d}}
\newcommand{\I}{\mathrm{i}}
\newcommand{\TP}[1]{{#1}^\mathrm{\,\textsc{t}}}
\newcommand{\E}[1]{\mathrm{e}^{\mbox{\footnotesize$#1$}}}

\newcommand{\HERM}[2]{\mathrm{H}_{\,#1}\!\left(#2\right)}
\newcommand{\KUMMERU}[3]{\mathrm{U}(#1,#2,#3)}

\AtBeginDocument{%
    \newwrite\bibnotes
    \def\bibnotesext{Notes.bib}
    \immediate\openout\bibnotes=\jobname\bibnotesext
    \immediate\write\bibnotes{@CONTROL{REVTEX41Control}}
    \immediate\write\bibnotes{@CONTROL{%
    apsrev41Control,author="08",editor="1",pages="1",title="0",year="1"}}
     \if@filesw
     \immediate\write\@auxout{\string\citation{apsrev41Control}}%
    \fi
}%

\begin{document}

\title{Benchmarking quantum tomography completeness and fidelity with machine learning}

\author{Yong~Siah~Teo}
\email{yong.siah.teo@gmail.com}
\affiliation{Department of Physics and Astronomy, 
	Seoul National University, 08826 Seoul, South Korea}

\author{Seongwook~Shin}
\affiliation{Department of Physics and Astronomy,
	Seoul National University, 08826 Seoul, South Korea}

\author{Hyunseok~Jeong}
\email{h.jeong37@gmail.com}
\affiliation{Department of Physics and Astronomy,
	Seoul National University, 08826 Seoul, South Korea}

\author{Yosep~Kim}
\affiliation{Department of Physics, Pohang University of Science and Technology (POSTECH), 37673 Pohang, Korea}

\author{Yoon-Ho~Kim}
\email{yoonho72@gmail.com}
\affiliation{Department of Physics, Pohang University of Science and Technology (POSTECH), 37673 Pohang, Korea}

\author{Gleb~I.~Struchalin}
\affiliation{Quantum Technologies Centre, and Faculty of Physics, Moscow State University, 119991 Moscow, Russia}

\author{Egor V. Kovlakov}
\affiliation{Quantum Technologies Centre, and Faculty of Physics, Moscow State University, 119991 Moscow, Russia}

\author{Stanislav S. Straupe}
\affiliation{Quantum Technologies Centre, and Faculty of Physics, Moscow State University, 119991 Moscow, Russia}

\author{Sergei P. Kulik}
\affiliation{Quantum Technologies Centre, and Faculty of Physics, Moscow State University, 119991 Moscow, Russia}

\author{Gerd~Leuchs}
\affiliation{Max-Planck-Institut f\"ur die Physik des Lichts, Staudtstra\ss e 2, 91058 Erlangen, Germany}
\affiliation{Institute of Applied Physics, Russian Academy of Sciences, 603950 Nizhny Novgorod, Russia}

\author{Luis~L.~S\'{a}nchez-Soto}
\email{lsanchez@fis.ucm.es}
\affiliation{Max-Planck-Institut f\"ur die Physik des Lichts,
	Staudtstra\ss e 2, 91058 Erlangen, Germany}
\affiliation{Departamento de \'Optica, Facultad de F\'{\i}sica,
	Universidad Complutense, 28040 Madrid, Spain}

\begin{abstract}
	We train convolutional neural networks to predict whether or not a set of measurements is informationally complete to uniquely reconstruct any given quantum state with no prior information. In addition, we perform fidelity benchmarking based on this measurement set without explicitly carrying out state tomography. The networks are trained to recognize the fidelity and a reliable measure for informational completeness. By gradually accumulating measurements and data, these trained convolutional networks can efficiently establish a compressive quantum-state characterization scheme by accelerating runtime computation and greatly reducing systematic drifts in experiments. We confirm the potential of this machine-learning approach by presenting experimental results for both spatial-mode and multiphoton systems of large dimensions. These predictions are further shown to improve when the networks are trained with additional bootstrapped training sets from real experimental data. Using a realistic beam-profile displacement error model for Hermite-Gaussian sources, we further demonstrate numerically that the orders-of-magnitude reduction in certification time with trained networks greatly increases the computation yield of a large-scale quantum processor using these sources, before state fidelity deteriorates significantly.
\end{abstract}

\maketitle

\section{Introduction}

Recent advances in quantum algorithms and error correction~\cite{Grimsley2019,Arute2019,Hu2019,Havlicek2019,Beer2020,Gard2020} have fueled the development of noisy intermediate-scale quantum computing devices. This progress requires an efficient assessment of the relevant quantum systems~\cite{Plesch:2011aa,Gard2020,Holmes:2020aa}, gates~\cite{Schafer:2018aa,Shi:2018aa,Ono:2017aa,Patel:2016aa,Fiurasek:2008fg} and measurements~\cite{Wootters:1989qf,NIELSEN:2003aa,Raussendorf:2001aa,Briegel:2009aa,Durt:2010cr,Scott:2006aa,Zhu:2011sp,Zhu:2014aa}. Toolkits developed in quantum tomography~\cite{Chuang:2000fk,lnp:2004uq,Teo:2015qs,OBrien:2004aa,Poyatos:1997aa,Teo2011aa,Luis:1999qm,Fiurasek:2001mq,D'Ariano:2004aa,Chen:2019aa,Zhang:2012aa,Altorio:2016aa} have concomitantly evolved into modern schemes appropriate for characterizing those components efficiently. A notable branch of schemes attempt to cope with a large number of qubits by directly estimating quantum properties~\cite{Kim:2018sw,Gaikwad:2018aa,Bendersky:2013aa,Schmiegelow:2011aa,Bendersky:2009aa,Bendersky:2008aa,Proctor:2017aa,Helsen2019,Yiping:2020aa}.

As typical quantum tasks involve pure states, unitary gates, and projective measurements, there also exists a series of compressed-sensing-related proposals~\cite{Gross:2010cs,Kalev:2015aa, Baldwin:2016cs,Steffens:2017cs,Schwemmer:2014aa,Riofrio:2017cs,Baldwin:2014aa,Rodionov:2014aa,Shabani:2011aa} that fully reconstruct low-rank quantum components with few measurement resources. However, they rely on prior knowledge about the rank, which often turns out to be unreliable in practice because of noise. Very recently, compressive-tomography methods without assuming any prior information has been developed and applied to the individual low-resource characterization of quantum states, processes and measurements~\cite{Ahn:2019aa,Ahn:2019ns,Teo:2020cs,Kim:2020aa,Gianani:2020aa}. A crucial ingredient in these methods is informational completeness certification that determines whether or not a given measurement set and its corresponding data is informationally complete~(IC). This is done by computing a uniqueness measure based on the given measurements. Such a computation can be performed with classical semidefinite programs~(SDPs)~\cite{Vandenberghe:1996ca} of (worst-case) polynomial-time complexities.

Like any tomography scheme that invokes rounds of optimization routines, an accumulation of errors can occur in real experiments while running SDPs on-the-fly. As a practically feasible solution, we propose to train an artificial neural network to verify the IC property for a set of quantum measurements and corresponding raw data. We further introduce an auxiliary network to be used concurrently for us to validate the fidelity of the unknown state for the given measurement set without carrying out explicit reconstruction. Once a set of IC measurement data is collected, it takes only one final round of state reconstruction to obtain the unique physical estimator, if so desired. 

Network training can be done offline using simulated noisy datasets, and the stored network model can later be retrieved and used in real experiments with statistical noise. More specifically, a convolutional neural-network~(CNN) architecture shall be used for training and prediction. Among other kinds of networks that have been widely adopted by the quantum-information community~\cite{Torlai2018,Palmieri2020,Neugebauer:2020aa,Lohani:2020aa,ahmed2020quantum,ahmed2020classification}, this is a popular network architecture that is used in image-pattern recognition~\cite{Lecun:1998aa,Krizhevsky:2012aa,Russakovsky:2015aa,He:2016aa}, with boosted support by a recent universality proof~\cite{Zhou:2020aa} that such networks can indeed forecast any continuous function mapping. Both its classical application and quantum analog have also gained traction in quantum-information science~\cite{Ming2019,Cong2019,Melnikov:2019aa,Tsai:2020aa}.

In this work, we train an \emph{informational completeness certification net}~(ICCNet) and a \emph{fidelity prediction net}~(FidNet), each made up of a sequence of convolution and pooling neural layers that is reasonably deep. Partnered with FidNet for direct fidelity benchmarking without the need for state tomography, ICCNet constitutes the foundational core for deciding if the given measurement resources are sufficient to uniquely characterize any unknown state in real experimental situations. Neural-network training is versatile in the sense that noise models may be incorporated into the training procedure to improve the predictive power of the networks. After offline training, the network models can heavily reduce the computation time of the uniqueness certification by orders of magnitude for large dimensions while running the experiments. This essentially realizes a compressive tomography scheme that is drift-proof, comprising a highly efficient uniqueness certification and fidelity-benchmarking protocols. 

Apart from Monte~Carlo simulations, we also use real data obtained from two separate groups of experiments to demonstrate that the resulting trained network models can predict the average behaviors of both the IC property and fidelity very well, despite the presence of errors and experimental imperfections. We also show that performances in predicting both properties can be further boosted when the neural networks are trained with additional bootstrapped experimental datasets. Finally, simulations on a time-dependent error model relevant to Hermite-Gaussian sources are performed as an example to illustrate the effectiveness of our neural-network certification tools in suppressing systematic drifts during quantum computation.

\begin{figure*}[t]
	\centering\includegraphics[width=2\columnwidth]{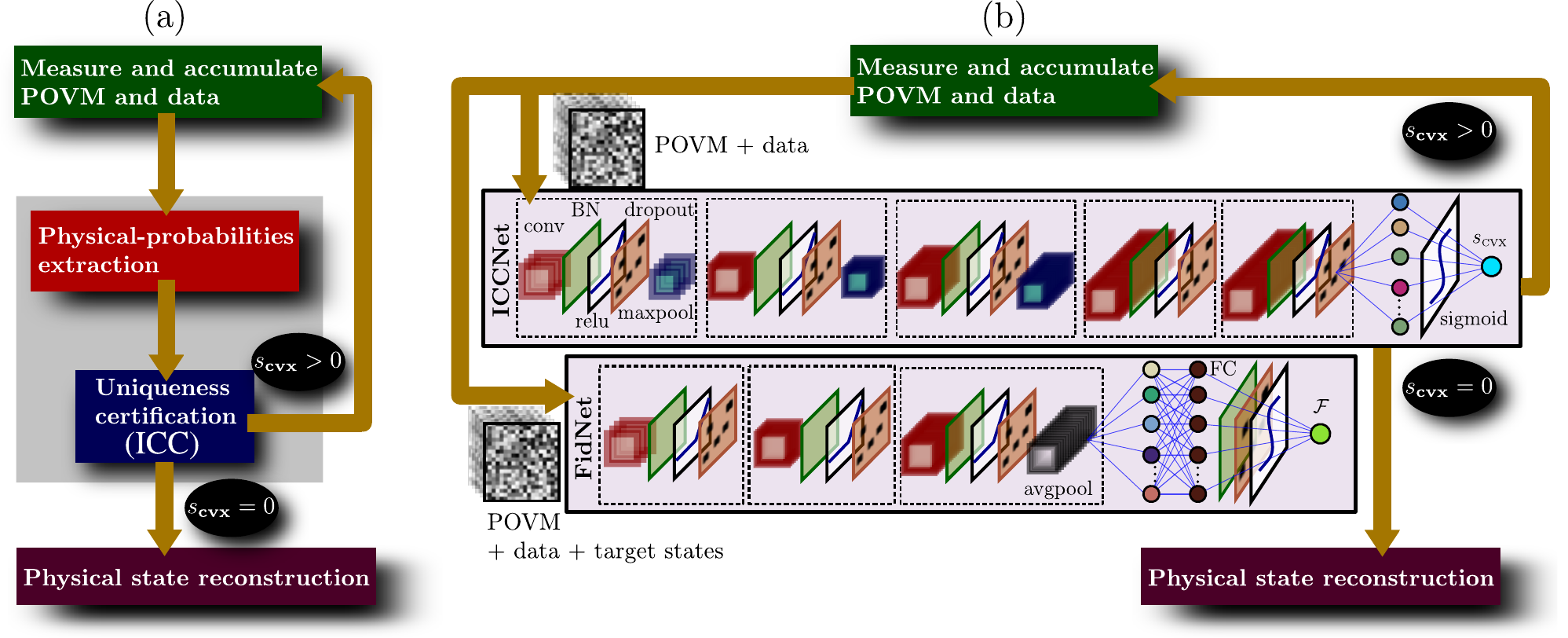}
	\caption{\label{fig:ICCNet_FidNet}The physical-probabilities extraction and SDP-based ICC of a resource-efficient quantum-state tomography scheme in (a) may be entirely replaced by the ICCNet and FidNet shown in (b), each of which is a sequence of convolutional blocks (shown here for $d=16$ as an example). Each convolutional block typically consists of a convolutional layer~(conv), a batch normalization layer~(BN), the relu activation layer, a dropout layer and a pooling layer~(maxpool or avgpool)~(more details in Sec.~\ref{subsec:net_training}). Specific network structures may vary for systems of different dimensions. Numerical values after the convolutional blocks are flattened and activated with the sigmoid function just before the $s_\textsc{cvx}$ computation, and passed through a fully-connected (FC) layer before the fidelity $\mathcal{F}$ computation.}
\end{figure*}

\section{Background}
\subsection{Ascertaining informational completeness}
\label{subsec:aIC}

The main procedure for certifying whether a generic measurement dataset is sufficient to unambiguously determine an unknown quantum state can be represented as a simple flowchart in Fig.~\ref{fig:ICCNet_FidNet}(a). Given a positive operator-valued measure (POVM) that models the measurements performed, the corresponding data counts are noisy due to statistical fluctuation arising from finite data samples. Proper data analysis first entails the extraction of physical probabilities from the accumulated data, which can be done with well-established statistical methods, such as those of maximum~likelihood~(ML)~\cite{lnp:2004uq,Rehacek:2007ml,Teo:2011me,Teo:2015qs,Shang:2017sf} and least~squares~(LS)~\cite{Kariya:2004aa,Rencher:2012aa}, subject to the physical constraint of density matrices (refer also to Sec.~\ref{subsec:aIC}).

After obtaining the physical probabilities, one may proceed to evaluate the measurements and find out whether they are IC. More specifically, a uniqueness indicator $0\leq s_\textsc{cvx}\leq1$ can be directly computed from the POVM and data with the help of SDPs---the \emph{informational completeness certification}~(ICC). When $s_\textsc{cvx}>0$, there is equivalently a convex set of state estimators that are consistent with the physical probabilities. It can be shown~\cite{Ahn:2019aa} that a unique estimator is obtained from the measured POVM and corresponding data if and only if $s_\textsc{cvx}=0$. 

Bottom-up resource-efficient quantum-state tomography is thus an iterative program involving rounds of extracting physical probabilities from the measurement data and certifying uniqueness based on these probabilities. At each round, the computed $s_\textsc{cvx}$ is used to decide whether new measurements are needed in the next one. In this manner, the POVM outcomes may be accumulated bottom-up until $s_\textsc{cvx} = 0$, after which a physical state reconstruction using either the ML or LS scheme is carried out to obtain the unique estimator. The size of the resulting IC POVM is minimized accordingly. We remark that ICC turns out to be the limiting procedure in practical implementation relative to a typical quantum-state reconstruction. This is because an estimation over the space of quantum states can be very efficiently implemented with an iterative scheme, where each step involves a regular gradient update and just one round of convex projection~\cite{Shang:2017sf}. (The case for quantum processes has also been discussed~\cite{Knee:2018aa}). On the other hand, satisfying both Born's rule and quantum positivity constraint in ICC requires a separate iterative procedure just to carry out the correct convex projection onto their intersection~\cite{Dykstra:1986aa}. To date, there exists no efficient way to perform projections of these constraints to the authors' knowledge.

By recalling the results in Refs.~\cite{Ahn:2019aa,Ahn:2019ns}, we briefly describe the simple procedures that deterministically verify whether a set of POVM outcomes $\{\Pi_j\geq0\}$ is IC given their corresponding set of relative frequency data $\{\nu_j\}$. The first necessary step is to acquire the physical probabilities from $\nu_j$. To this end, we consider two popular choices often considered in quantum tomography, namely the ML and LS methods. In ML, we maximize the log-likelihood function $\log L$ that best describes the physical scenario over the quantum state space. Since we predominantly discuss von~Neumann measurement bases, each basis induces a multinomial distribution such that we have the form $\log L\propto\sum_j\nu_j\log p'_j$, where $p'_j=\tr{\rho'\Pi_j}$ are our sought-after physical probabilities to be optimized over the operator space in which $\rho'\geq0$ and $\tr{\rho'}=1$. In LS, which we have adopted to deal with arbitrary projective measurements that do not sum to the identity operator in general, the distance $\mathcal{D}=\|\nu_j-p'_j\|^2$ is minimized with respect to $p'_j$ over the space of $\rho'\geq0$, this time with the unit-trace constraint relaxed and later reinstated after the minimization is completed.

Using the obtained physical probability estimators $\widehat{p}_j$ through the aforementioned optimization strategies, we can now define and fix a randomly generated full-rank state $Z$ and conduct the following two SDPs:
\begin{align}
	&\,\mathrm{minimize/maximize}~f=\tr{\rho'Z}\,\text{over }\rho'\nonumber\\
	&\,\text{subject to:}\,\nonumber\\
	&\,\rho'\geq 0\,,\quad\tr{\rho'}=1\,,\quad
	\tr{\rho'\Pi_j}=\widehat{p}_j\,.
	\label{eq:SDP}
\end{align}
It is now clear why the SDPs are to be carried out with the physical probabilities $\widehat{p}_j$ instead of raw data $\nu_j$: the relative frequencies $\nu_j$ are statistically noisy and in general do not correspond to a feasible solution set in \eqref{eq:SDP}. It has been shown that when $s_\textsc{cvx}\equiv f_\mathrm{max}-f_\mathrm{min}$ is zero, this implies that any quantum-state estimator reconstructed from $\{\Pi_j\}$ and $\{\nu_j\}$ is unique and equal to the solution for \eqref{eq:SDP}. 

\begin{figure}[t]
	\centering\includegraphics[width=1\columnwidth]{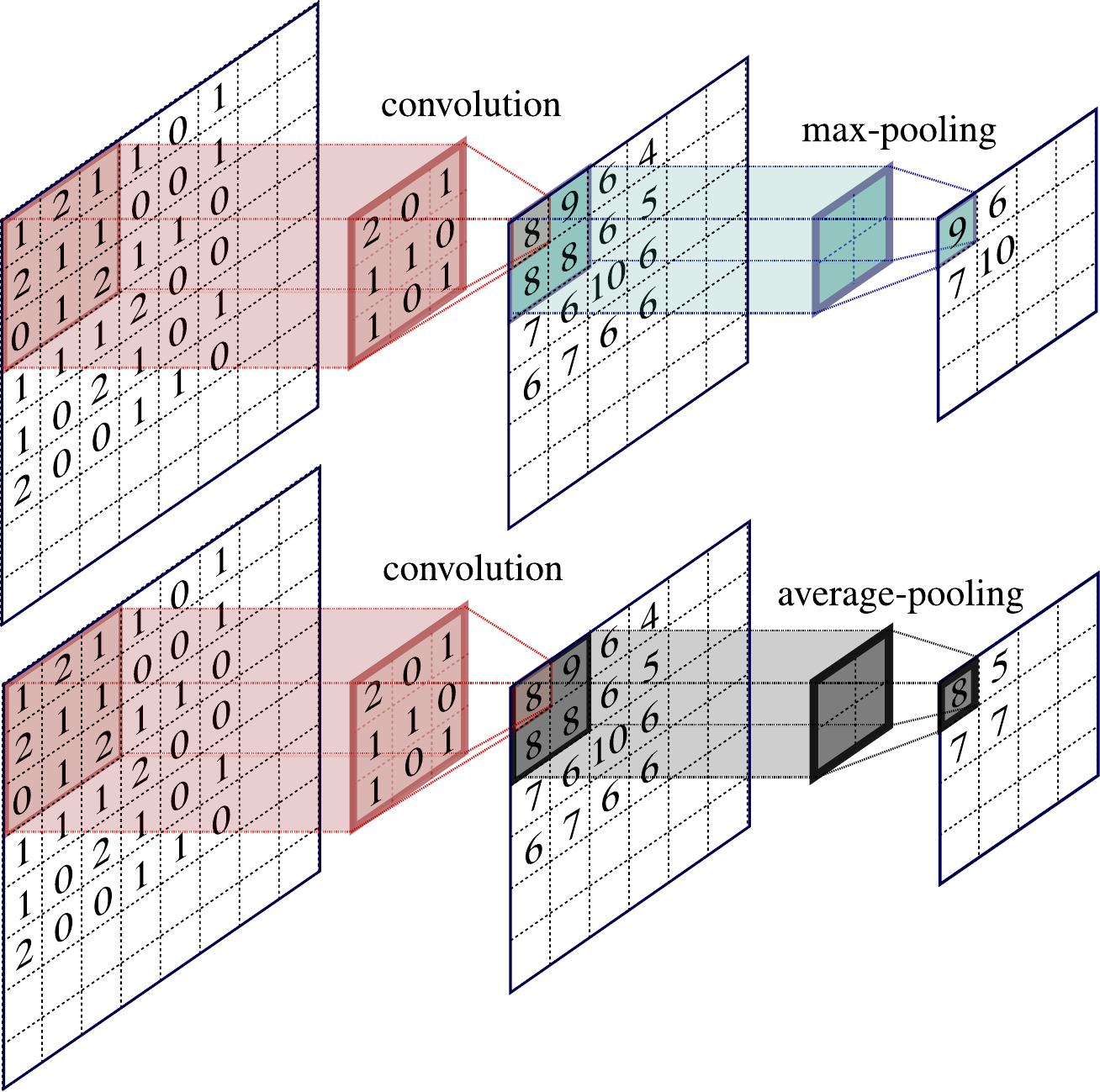}
	\caption{\label{fig:conv_max_avg}Operations carried out by the convolution and pooling layers. A max-pooling layer picks the maximum number from the layer input within a selected window, while an average-pooling layer computes average values over the selected window. In this example, the $8\times 8$ input layer is reduced to a $6\times6$ output layer after going through a convolution layer consisting of a single $3\times3$ filter array of trainable parameters that takes stride~1. This output layer becomes the input layer with respect to either the max-pooling or average-pooling layer that each consists of a single $2\times2$ filter array of stride~2. The final output layer (rounded off for illustration purposes) is therefore a $4\times4$ numerical array.}
\end{figure}

\subsection{Training the ICCNet and FidNet}
\label{subsec:net_training}

We propose to tackle the combined problem of physical probabilities extraction and uniqueness certification by predicting with trained neural networks. We also demonstrate the possibility of performing fidelity evaluation on the reconstruction with such neural networks without explicitly carrying out physical state tomography. To do this, we introduce the ICCNet and FidNet, illustrated in Fig.~\ref{fig:ICCNet_FidNet}(b), where each possesses a convolutional network architecture that analyzes the given POVM and data by regarding them as images. Such a treatment allows one to train the networks with far less trainable parameters to recognize $s_\textsc{cvx}$ and fidelity $\mathcal{F}(\widehat{\rho},\rho_\mathrm{targ})=\tr{\sqrt{\sqrt{\widehat{\rho}}\,\rho_\mathrm{targ}\sqrt{\widehat{\rho}}}}^2$ between the state estimator $\widehat{\rho}$ and some target state $\rho_\mathrm{targ}$ as compared to using, for instance, the multilayer perceptron (feed-forward) architecture~\cite{Rumelhart:1987aa,Hastie:2009aa} that consists only of fully-connected or dense layers. 

The purpose of FidNet is to assess the quality of the reconstruction after each measurement set is made. Before the point of informational completeness, the reconstructed state $\widehat{\rho}$ is not unique by definition. Throughout this article, for consistency, $\widehat{\rho}$ shall always be taken as the ML estimator that minimizes the linear function in \eqref{eq:SDP}. This is only a particular choice used to define FidNet that is chosen as a standard. One may end up with a slightly more conservative FidNet by setting $\widehat{\rho}$ to be the minimum-fidelity estimator with respect to the SDP constraints stated in the last line of \eqref{eq:SDP}. This would require another round of fidelity minimization in every step of the training-data-generation phase.

For predicting $s_\textsc{cvx}$ and $\mathcal{F}$, both ICCNet and FidNet employ a sequence of two-dimensional array manipulating layers. Two important types of layers responsible for these operations are the convolution layer, which are two-dimensional filters that carry out multiplicative convolutions with the layer input numerical array, and the pooling layer that generally down-samples a layer input array into a smaller output array with a simple numerical-summarizing computation. To each convolution layer, an activation function is applied to further introduce nonlinear characteristics for predicting general network output functions.

The convolutional ICCNet and FidNet take on a similar architecture, which consists of convolution, max-pooling and average-pooling layers. Each convolution layer consists of $n_\mathrm{f}$ filters, where each filter is a $3\times3$ array window that slides vertically and across layer input arrays with stride 1 in both directions. We design the sequence of convolution layers to have an exponentially increasing $n_\mathrm{f}$ with the network depth. These pooling layers are generally responsible for shrinking the layer input array to a smaller layer output array. The actions of all types of layers are summarized in Fig.~\ref{fig:conv_max_avg}. We insert the default rectified linear unit (``relu'') activation function after every convolution layer, which is defined as $f_\mathrm{relu}(x)=\max(0,x)$. At the end of ICCNet and FidNet, the respective output values are computed with the sigmoid activation function given by $f_\mathrm{sigmoid}(x)=1/(1+\E{-x})$.

Overfitting can be an issue in machine learning, in which case the neural networks are prone to fitting training datasets much better than unseen ones. It is therefore essential to regulate network training by keeping the problem of overfitting in check so that the resulting trained models have high predictive power. This problem often arises when the neural network is deep. The addition of dropout layers has been proven to be an effective method for combating overfitting~\cite{Hinton:2012aa,Srivastava:2014aa,Warde-Farley:2014aa}, which randomly exclude trainable parameters. More recently, it has been demonstrated that neural-network training can be further enhanced by adding batch normalization layers. This was supported not only by the initial observation that the distribution of layer input values are stabilized with batch normalization~\cite{Ioffe:2015aa}, but also by the even more relevant finding that the gradient landscape of the network loss function (the figure of merit quantifying the difference between the actual output value and that computed by the network) seen by the optimization routine that trains the network becomes smoother~\cite{Santurkar:2018aa}, making training much more stable.

\begin{figure}[t]
	\centering\includegraphics[width=1\columnwidth]{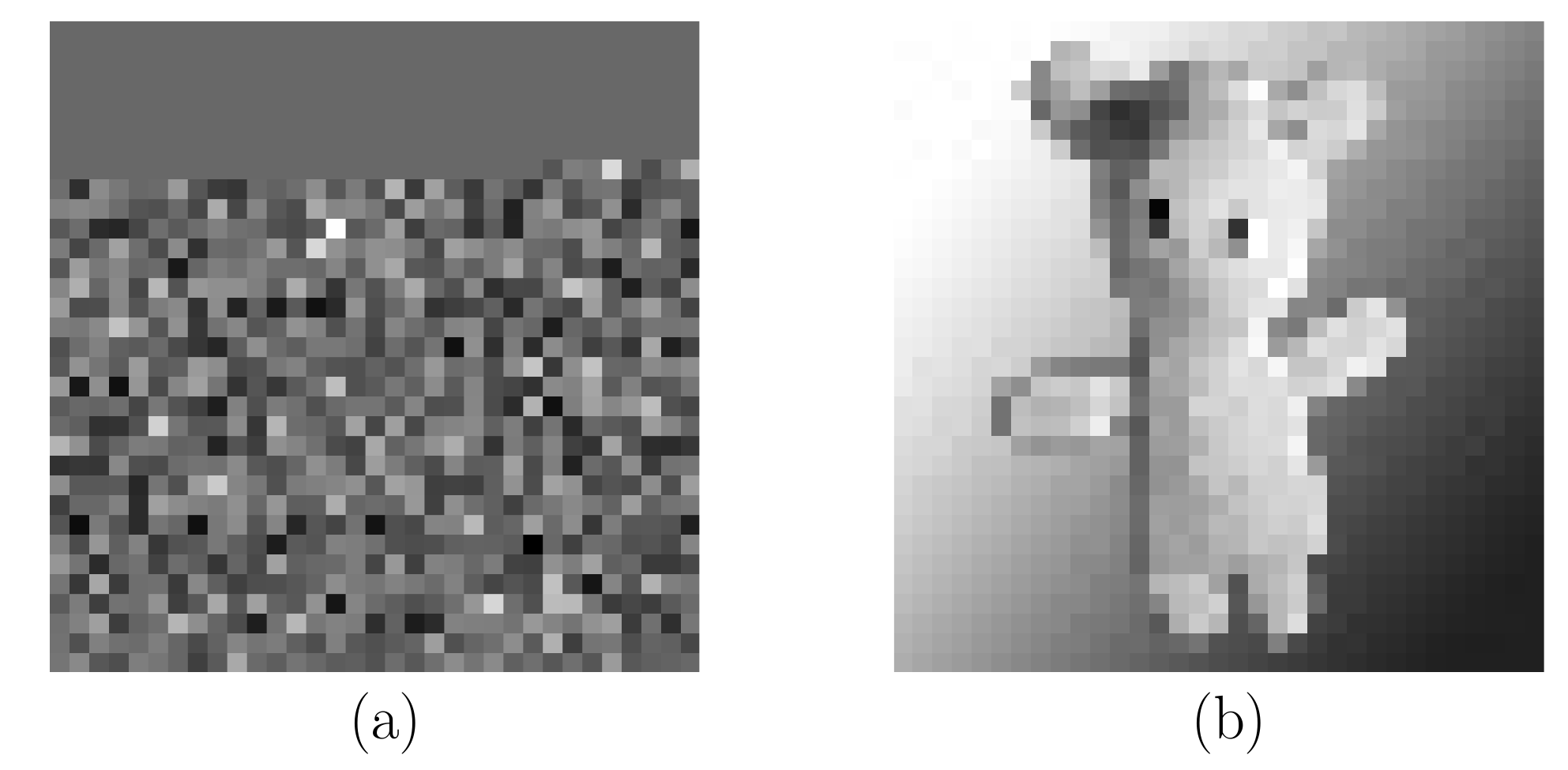}
	\caption{\label{fig:qimg_vs_pig}A juxtaposition of (a)~a $33\times33$ pixelated ICCNet input-data image, which encodes a four-qubit POVM containing $K=4$ bases and corresponding probabilities, and (b)~a down-sampled photograph of a stuffed toy of the same resolution. Here, the ICCNet input-data image is generated by proportionately scaling all numerical values in the square training array $\dyadic{X}$ to values between 0 and 255 only for the purpose of illustrative comparison.}
\end{figure}

All trainable parameters in the relevant neural layers of \mbox{ICCNet} and FidNet are optimized using a variant of stochastic gradient descent known as NAdam~\cite{Dozat:2016aa}, where the network gradients are computed in batches of the training data. To prepare ICCNet training input datasets, for von~Neumann measurements of a fixed number ($K$) of bases considered in Secs.~\ref{subsec:sim} and \ref{subsec:expt1}, the initial network input $\dyadic{X}$ is an $m\times K(d^2+d)$ matrix that contains $m$ training datasets, each recording the $K$ measured bases and corresponding relative frequencies $\{\nu_{jk}\}^{d-1}_{j=0}\,{}^K_{k=1}$ ($\sum_j\nu_{jk}=1$). To encode the measurement bases, we regard all bases as some unitary rotation $U_k\ket{j}\bra{j}U_k^\dag$ of the standard computational basis $\{\ket{j}\}^{d-1}_{j=0}$, where $U_1=1$. These unitary operators are then logarithmized in order to obtain their Hermitian exponents $H_k=-\I\log U_k$ ($H_1=0$), from which the diagonals and upper triangular real and imaginary matrix elements are extracted. Each row of $\dyadic{X}$ is thus a flattened $K(d^2+d)$-dimensional row of real numerical values formatted properly to encode $U_1$, $U_2$, $\ldots$, $U_K$, $\nu_{01}$, $\ldots$, $\nu_{d-1\,1}$, $\nu_{02}$, $\ldots$, $\nu_{d-1\,2}$, $\ldots$, $\nu_{0\,K}$, $\ldots$, $\nu_{d-1\,K}$ in this order. This input matrix is then processed into a $\lceil\sqrt{K(d^2+d)}\rceil\times\lceil\sqrt{K(d^2+d)}\rceil$ square training array of elements, which is then fed to ICCNet~(see Fig.~\ref{fig:qimg_vs_pig}). Zeros are padded to this array in order to complete the square. Similarly, for a fixed set of $L$ projective measurements discussed in Sec.~\ref{subsec:expt2}, analogous arguments lead to the necessary $\lceil\sqrt{L(d^2+1)}\rceil\times\lceil\sqrt{L(d^2+1)}\rceil$ input square training array. For each dimension, the randomly generated full-rank state $Z$ needed to solve \eqref{eq:SDP} is fixed during the training and testing stages.

On the other hand, training the FidNet requires input information about not only the measured bases (or projectors) and their corresponding data, but also the additional $m$ target states to be included as inputs, one for each dataset. The correct dimensions of the training arrays are $\lceil \sqrt{(K+1)d^2+Kd}\rceil \times \lceil \sqrt{(K+1)d^2+Kd}\rceil$ or $\lceil\sqrt{(L+1)d^2+L}\rceil\times\lceil\sqrt{(L+1)d^2+L}\rceil$ respectively for basis and projective measurements. We note that to predict fidelities for simulated test datasets of $d=16$, 32 and 64 as shown in Fig.~\ref{fig:NNHaarACT_d16} and \ref{fig:time_graphs}, 

\emph{For all purely-simulation figures,} FidNet training is done with target states defined by the true states that generated the simulated training datasets. On the other hand, for all experimental results in Fig.~\ref{fig:ICCNN_FidNN_expt}, FidNet training is carried out simultaneously with the target states derived from the corresponding true states and those that deviate from them in order to account for systematic errors more effectively and improve average prediction accuracy. The list of hyperparameters that define the architectures of ICCNet and FidNet, as well as the technical analyses of network input-data generation and network training are given in Appendices~\ref{app:hyper} and \ref{app:explicit}.

Once an IC set of measurements are performed and assessed with ICCNet and FidNet, the density matrix representing the final state estimator may be obtained using the accelerated projected-gradient algorithm developed in \cite{Shang:2017sf}. Alternatively, it is possible to append our networks with additional conditional generative networks to yield the density matrix~\cite{ahmed2020quantum,ahmed2020classification}.

\section{Experiments}

\begin{figure}[t]
	\centering
	\includegraphics[width=3.4in]{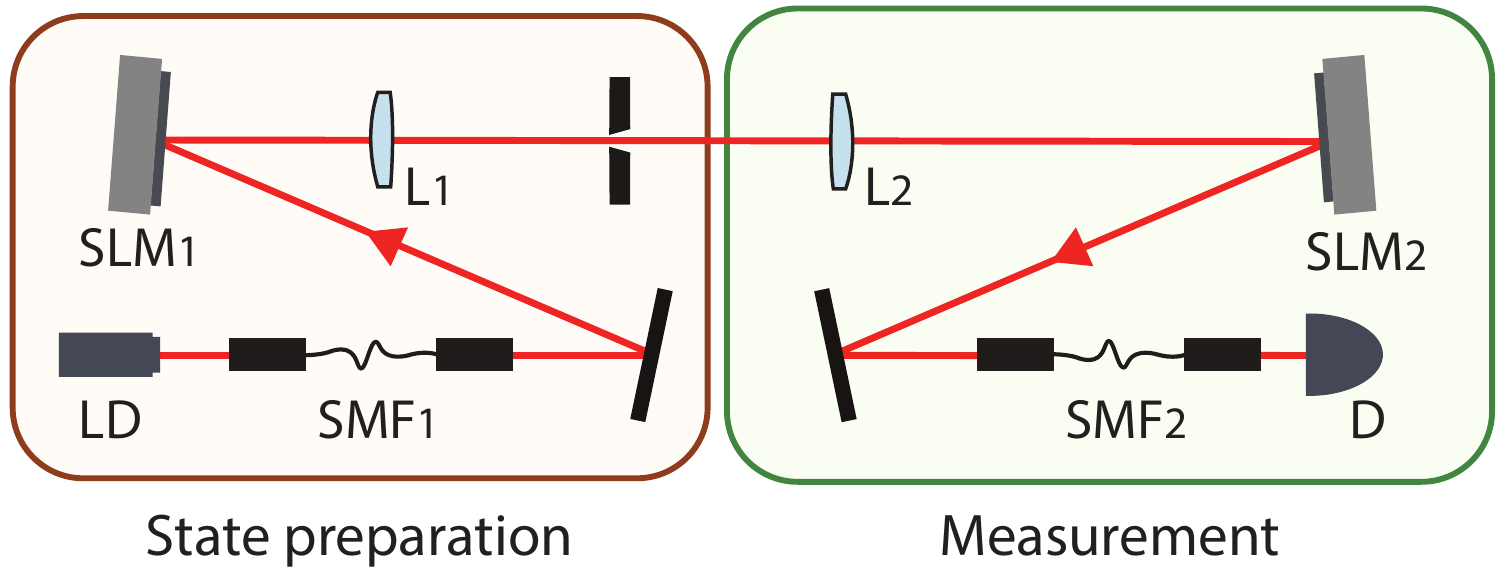}
	\caption{Experimental scheme to generate and characterize spatial photon states. Attenuated radiation of laser diode (LD) is spatially filtered by a single-mode optical fiber (SMF1) and directed on the first spatial light modulator (SLM1). Hologram displayed on the SLM1 transforms the fundamental fiber mode into the desired superposition of Hermite-Gaussian beams defining the particular quantum state of photons. The iris placed in the middle of the telescope with unit magnification (lenses L1 and L2) is used to clean the structured beam from the undiffracted light by selection of the first order of diffraction at the far-field plane of the  SLM1. The second light modulator (SLM2) followed by a single-mode optical fiber (SMF2) and a single photon counter (D) plays a role of spatial detector, which realize a projective measurement by the right choice of a hologram on the second SLM display.
	}\label{fig:qspace}
\end{figure}

\subsection{Spatial-mode photonic systems}

Apart from evaluating simulation test datasets, we also run the trained ICCNet and FidNet models to benchmark real experimental datasets. In the first group of experiments, we showcase the accuracy of ICCNet and FidNet predictions on experimental data acquired from an attenuated laser source prepared in quantum states projected onto Hermite-Gaussian spatial-mode bases of various dimensions $d$. With this group of experiments, for the sake of variety, we shall consider measurement bases that are obtained from adaptive compressive tomography~(ACT). These are eigenbases of the state that minimizes the von~Neumann entropy subject to the same SDP constraints in \eqref{eq:SDP}. It has been demonstrated that successive measurements of such eigenbases result in a fast convergence of $s_\textsc{cvx}$~\cite{Ahn:2019aa,Ahn:2019ns}. An explicit protocol to construct these bases is given in Appendix~\ref{app:gen_rand_meas}. 

The Hilbert space of photonic spatial degrees of freedom is typically discretized using an appropriate basis of transverse modes. To produce high-dimensional quantum states we attenuate an 808-nm diode laser, filter the resulting radiation with a single-mode optical fiber and then adjust the spatial structure of the light field with a spatial light modulator (SLM, see Fig.~\ref{fig:qspace}). The holographic approach~\cite{bolduc2013exact} allows us to transform the incident light into arbitrary transverse modes by controlling the phase pattern on the SLM's display. 

We work with Hermite-Gaussian~(HG) modes $\mathrm{HG}_{nm}(x,y)$, which are the solutions to the Helmholtz equation in Cartesian coordinates $(x,y)$ and form a complete orthonormal basis. By bounding the sum of beam orders $n+m$, we restrict the dimension of the generated quantum systems. Since holograms displayed on the SLM make use of a blazed grating, in order to select the first diffraction order, we place an iris in the middle of the telescope, where different diffraction orders are well separated. Using a second SLM, a single-mode optical fiber, followed by a single photon counting module, we realize a well-known technique of projective measurements in the spatial-mode space~\cite{mair2001entanglement}. These allows us to also conveniently implement general ACT basis measurements in arbitrary dimensions.

\begin{figure}[t]
	\centering
	\includegraphics[width=3.4in]{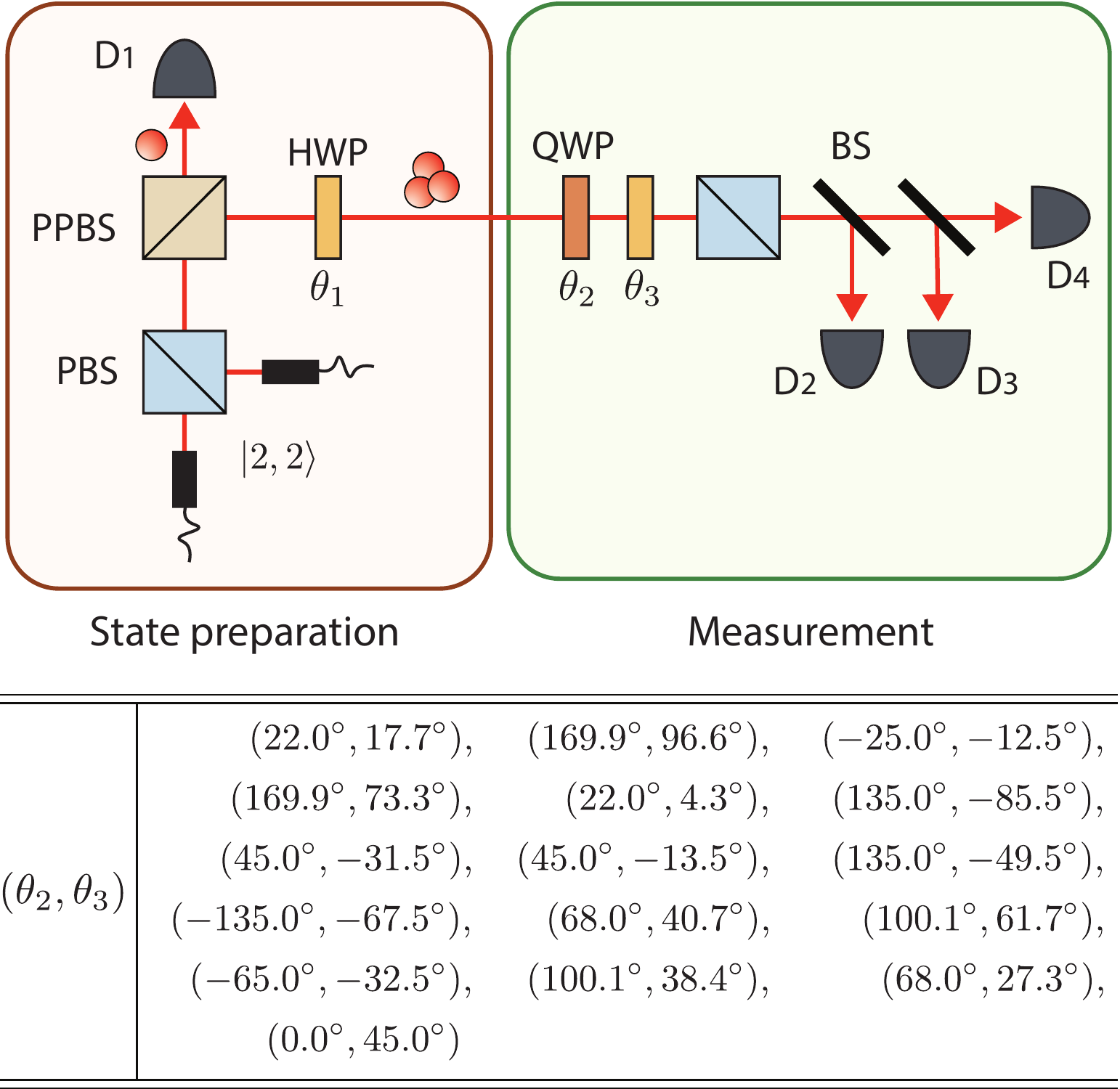}
	\caption{Experimental scheme to generate and characterize three-photon states. Two horizontally polarized photons and two vertically polarized photons, produced by the double-pair emission of non-collinear spontaneous parametric down-conversion process, are spatially combined with a PBS, thereby producing the four-photon state $\ket{2,2}\bra{2,2}$. After detecting a single photon at detector D$_1$, the reflected three-photon system from a PPBS are prepared in a particular quantum state, determined by the HWP angle $\theta_1$. For state characterization, four-fold coincidence counts at detectors D$_1$, D$_2$, D$_3$, and D$_4$ are acquired for all 16 rank-one projectors pictorialized in Fig.~\ref{fig:ICCNN_expt2_meas}. These measurement projectors are determined by the HWP and QWP angles of $\theta_2$ and $\theta_3$ in the table with a PBS and BS.
	}\label{fig:qpol}
\end{figure}

\subsection{Multiphoton systems}
\label{subsec:multiphoton}

In the second group of experiments, we switch to a different flavor of informational completeness by discussing two-mode photon-number states. In particular, we look at quantum states of up to three photons occupying two optical modes. Such three-photon states were of interest in the study of high-order quantum polarization properties beyond the Stokes vectors~\cite{Kim:2017aa}. The resulting Hilbert space is effectively 4-dimensional and spanned by the basis $\{\ket{n_{\textsc{h}},n_{\textsc{v}}}\}_{n_{\textsc{h}}+n_{\textsc{v}}=3}=\{\ket{0,3},\ket{1,2},\ket{2,1},\ket{3,0}\}$. Here $n_{\textsc{h}}$ and $n_{\textsc{v}}$ denote the number of photons in the horizontal and vertical polarization modes, respectively.

To perform tomography on the multiphoton quantum states, expectation values of a set of 16 rank-one projectors are measured. In principle, any set of 16 linearly independent projectors are suitable for a complete characterization of arbitrary 4-dimensional states \emph{without} ICC. For these experiments, we define each projector $\Pi_j$ by a ket ${b_j^{\dagger}}^3\ket{0,0}/\sqrt{6}$, where $b_j^{\dagger}$ and the other unobserved counterpart $c_j^{\dagger}$ are photonic creation operators derived from an SU(2) unitary operator $\widetilde{U}_j$ according to the transformation
\begin{equation}
	\begin{pmatrix}
		b^{\dagger}_j \\[1ex]
		c^{\dagger}_j
	\end{pmatrix}=\widetilde{U}_j\begin{pmatrix}
		a_\textsc{h}^{\dagger} \\ 
		a_\textsc{v}^{\dagger} 
	\end{pmatrix}\,,
	\label{eq:transf}
\end{equation}
and $a^{\dagger}_\textsc{h}$ and $a^{\dagger}_\textsc{v}$ are the creation operators of the horizontal and vertical polarization modes~\cite{Schilling:2010aa}. Clearly, $\sum_j\Pi_j\neq1$ this time, as the projectors are independently measured.

\begin{figure}[t]
	\centering\includegraphics[width=0.6\columnwidth]{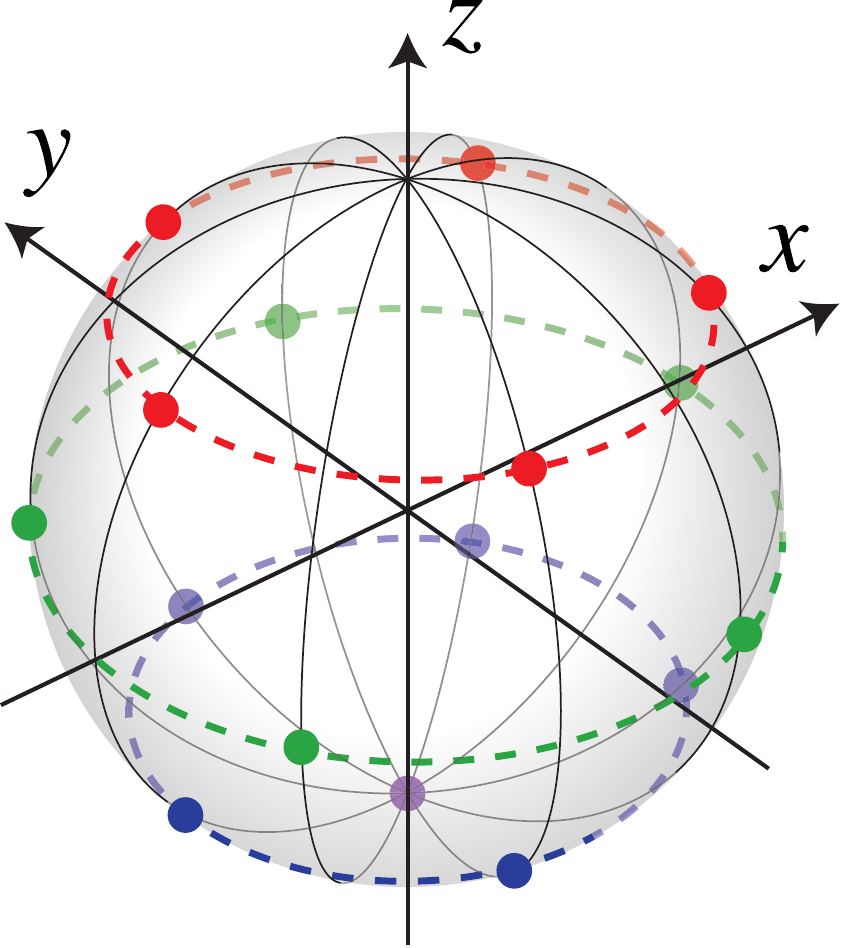}
	\caption{\label{fig:ICCNN_expt2_meas} Reduced visualization of the 16 two-photon measurement projectors on the single-qubit Bloch sphere. The projectors of three-photon states are defined as ${b_j^{\dagger}}^3\ket{0,0}/\sqrt{6}$ in accordance with Eq.~\eqref{eq:transf}. The projection states are chosen to equally distribute the corresponding single-photon component pure states $b_j^{\dagger}\ket{0,0}$ on the equatorial great circle and two small circles on the Bloch sphere, together with the south pole.}
\end{figure}

Figure~\ref{fig:qpol} depicts the experimental setup to generate and characterize three-photon states. Four photons are produced through double pair emission of non-collinear spontaneous parametric down-conversion (SPDC) process. The initial state is prepared in $|2,2\rangle$ by combining two horizontally polarized photons and two vertically polarized photons with a polarizing beam splitter~(PBS). To ensure that the photons are indistinguishable in the frequency domain, interference filters of 3~nm bandwidth centered at 780~nm are placed before sending the photons into the PBS. The four photons are then reduced into three photons by detecting a photon at $\text{D}_1$ and the reflected three photons from a partially-polarizing beam splitter~(PPBS) are in the state of $\ket{1,2}\bra{1,2}$. The PPBS perfectly reflects vertically polarized photons and reflects 1/3 of horizontally polarized photons. The half-wave plate (HWP) setting of $\theta_1=0^{\circ}$ leaves the state unchanged, whereas the setting of $\theta_1=45^{\circ}$ transforms the state into $\ket{2,1}\bra{2,1}$. In addition, the mixed state $(\ket{1,2}\bra{1,2}+\ket{2,1}\bra{2,1})/2$ is obtained by incoherently adding the relevant pure states through post-processing. These three-photon states are used to demonstrate the performances of ICCNet and FidNet in Fig.~\ref{fig:ICCNN_FidNN_expt}(b).

Experimentally~\cite{Kim:2017aa}, the three-photon states were characterized by acquiring the four-fold coincidence counts at D$_1$, D$_2$, D$_3$, and D$_4$ for 16 rank-one projectors after passing through a PBS and beam splitters (BS). The SU(2) unitary operators $\widetilde{U}_j$ that define the projectors $\Pi_j={b_j^{\dagger}}^3\ket{0,0}\frac{1}{6}\bra{0,0}b_j^3$ according to rule~\eqref{eq:transf} are determined by the quarter-wave plate (QWP) and HWP angles of $\theta_2$ and $\theta_3$ inasmuch as $\widetilde{U}_j=H(\theta_3)Q(\theta_2)$, where the matrix representations for the wave plates are given by
\begin{align}
	Q(\theta)\,\widehat{=}&\,\frac{1}{\sqrt{2}}\begin{pmatrix}
		1-\I\cos(2\theta) & -\I\sin(2\theta) \\
		-\I\sin(2\theta) & 1+\I\cos(2\theta) 
	\end{pmatrix}, \nonumber\\
	H(\theta)\,\widehat{=}&\,\begin{pmatrix}
		\cos(2\theta) & \sin(2\theta) \\
		\sin(2\theta) & -\cos(2\theta) \\
	\end{pmatrix}.
\end{align}
In our experiments, we consider SU(2) rotations that fairly distribute the single-photon component $b_j^{\dagger}|0,0\rangle$ on three Bloch-spherical circles parallel to the equatorial plane~\cite{Kim:2017aa,Israel:2012aa} as shown in Fig.~\ref{fig:ICCNN_expt2_meas}. The measurement angles that realize these projectors are given in Fig.~\ref{fig:qpol}.

\section{Results}
\label{sec:res}

\begin{figure}[t]
	\centering\includegraphics[width=1\columnwidth]{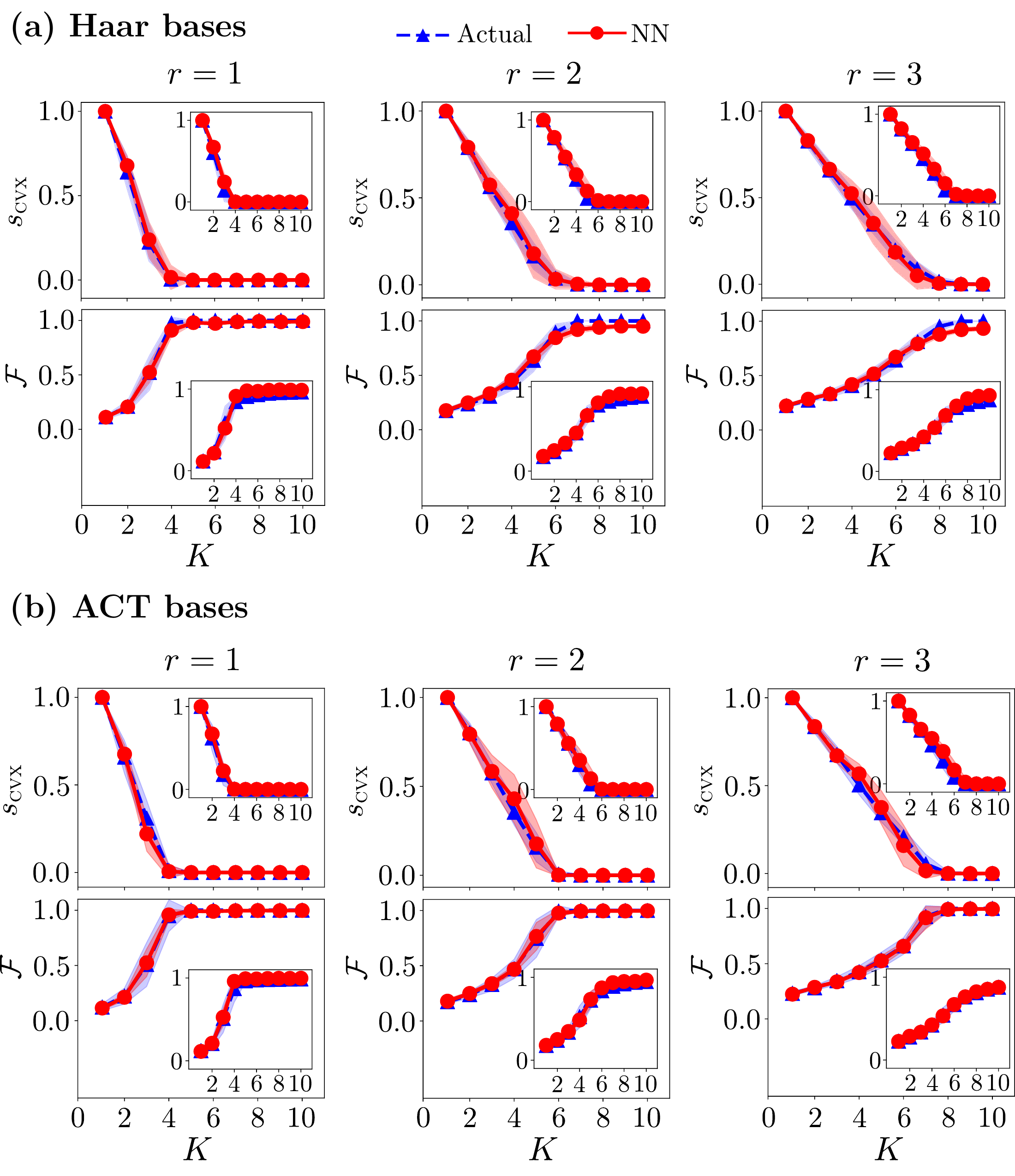}
	\caption{\label{fig:NNHaarACT_d16}Performance of ICCNet and FidNet in the prediction of $s_\textsc{cvx}$ and $\mathcal{F}$ for different number~($K$) of measurement bases generated by (a)~random unitaries sampled from the Haar unitary group and (b)~adaptive unitaries from ACT, accompanied by 1-$\sigma$ error bars derived from 50 simulated test experiments for each rank $r$ that are not seen by the neural networks. The main plots correspond to perfect measurement data, whereas the insets show results under statistical noise with $N=1000$ sampling copies per basis. Both the actual computed values and neural-network~(NN) predictions are evidently in extremely good agreement.}
\end{figure}

\begin{figure}[t]
	\centering\includegraphics[width=0.8\columnwidth]{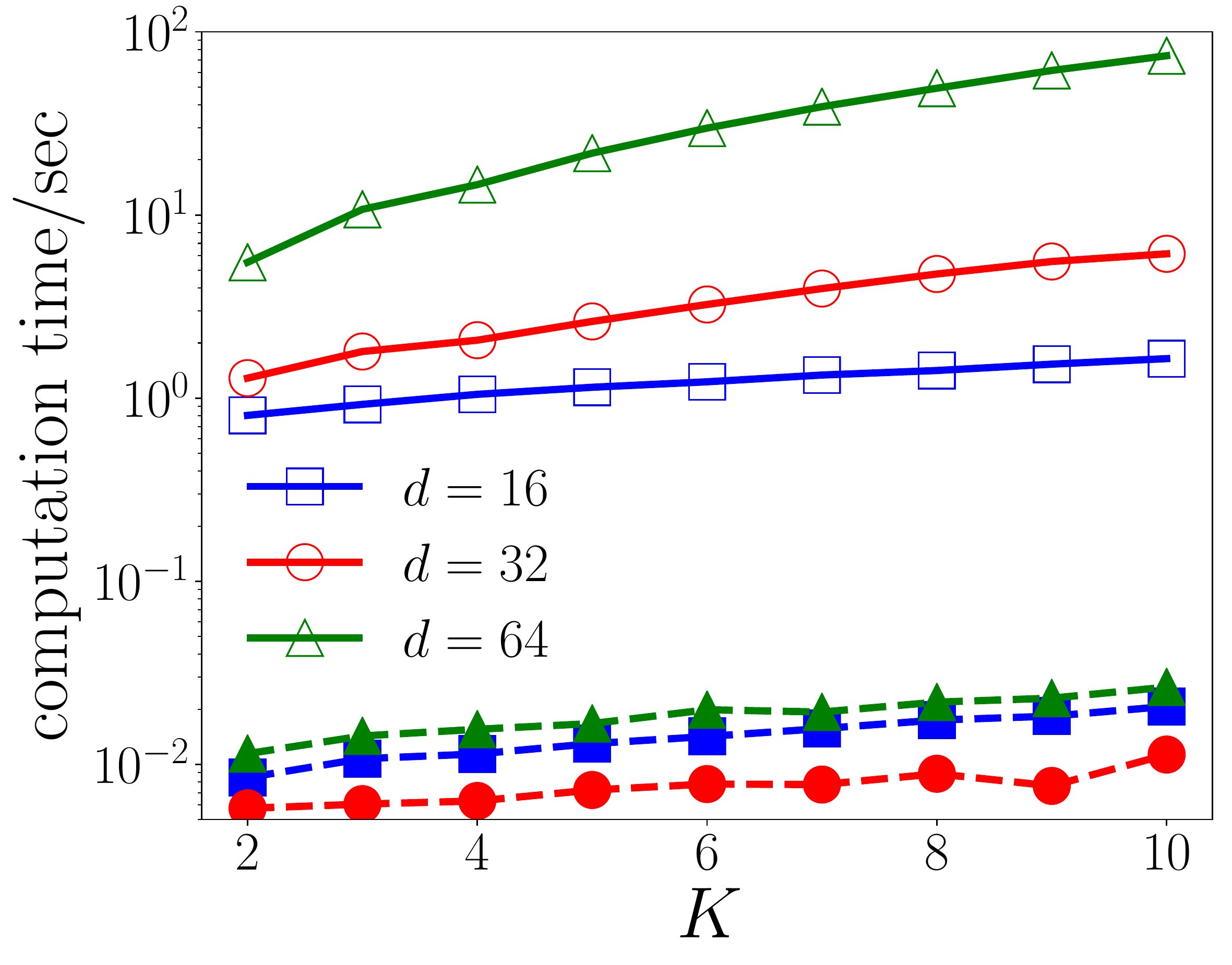}
	\caption{\label{fig:time_graphs}Comparison of the average ICC computation time by carrying out the grayed subroutine (physical-probabilities extraction and SDP-based ICC) in Fig.~\ref{fig:ICCNet_FidNet} (unfilled markers) and a trained ICCNet model (solid markers) over many simulated experimental runs and states of various ranks. For $d=32$ and 64, a set of 1000 datasets ($N=1000$) each is used to acquire average computation times that are sufficiently representative (the $s_\textsc{cvx}$ and $\mathcal{F}$ graphs are separately presented in Fig.~\ref{fig:scvx_fid_d32_d64}). These timing are obtained through CUDA~10.2 interfaced with the GPU-enabled TensorFlow~1.9 package on Python 3.5.3, with the Keras~2.1.6 frontend running on a twelve-core Intel(R) Xeon(R) CPU E5-2620~v3 at 2.40~GHz and an Nvidia GTX~1080~TI GPU of native settings. A trained FidNet model, on average, performs fidelity benchmarking in times that are roughly the same orders of magnitude. That the $d = 16$ neural-network time curve is between those for $d = 32$ and 64 is due to neural-network-architectural differences for different $d$ values. Performance gaps are barely noticeable in practice.}
\end{figure}

\subsection{Simulations --- Neural-network performances}
\label{subsec:sim}

We first present performance graphs of ICCNet and FidNet in Fig.~\ref{fig:NNHaarACT_d16} based on two sets of simulations on four-qubit states~($d=16$) using random measurement bases generated with the Haar measure for the unitary group~(see Appendix~\ref{app:gen_rand_meas}), and bases found using ACT. In each set of simulations, for both cases where statistical noise is either absent or present, we collect simulation data of various number ($K$) of bases ($s_\textsc{cvx}$ is normalized to 1 at $K=1$ by default), each case recording measurements of 5000 randomly-generated quantum states of uniformly distributed rank $1\leq r\leq 3$. The explicit CNN architecture employed is specified in Sec.~\ref{subsec:net_training}. The accurate fit between the actual computed values and those predicted by ICCNet and FidNet suggests that faithful neural network predictions of both the degree of informational completeness and fidelity are a definite possibility in both noiseless and statistically noisy environments. Sample codes for network training and evaluation with four-qubit simulation datasets are available online~\cite{ICC_FID_github:2021aa}.

In separate simulations on four- ($d=16$), five- ($d=32$) and six-qubit ($d=64$) systems with random Haar measurement bases, numerical evidence presented in Fig.~\ref{fig:time_graphs} shows that the computation times in $s_\textsc{cvx}$ neural-network predictions can be significantly reduced by about four orders of magnitude relative to ordinary SDP calculations, and this difference grows wider with larger dimensions.

\begin{figure}[t]
	\centering\includegraphics[width=1\columnwidth]{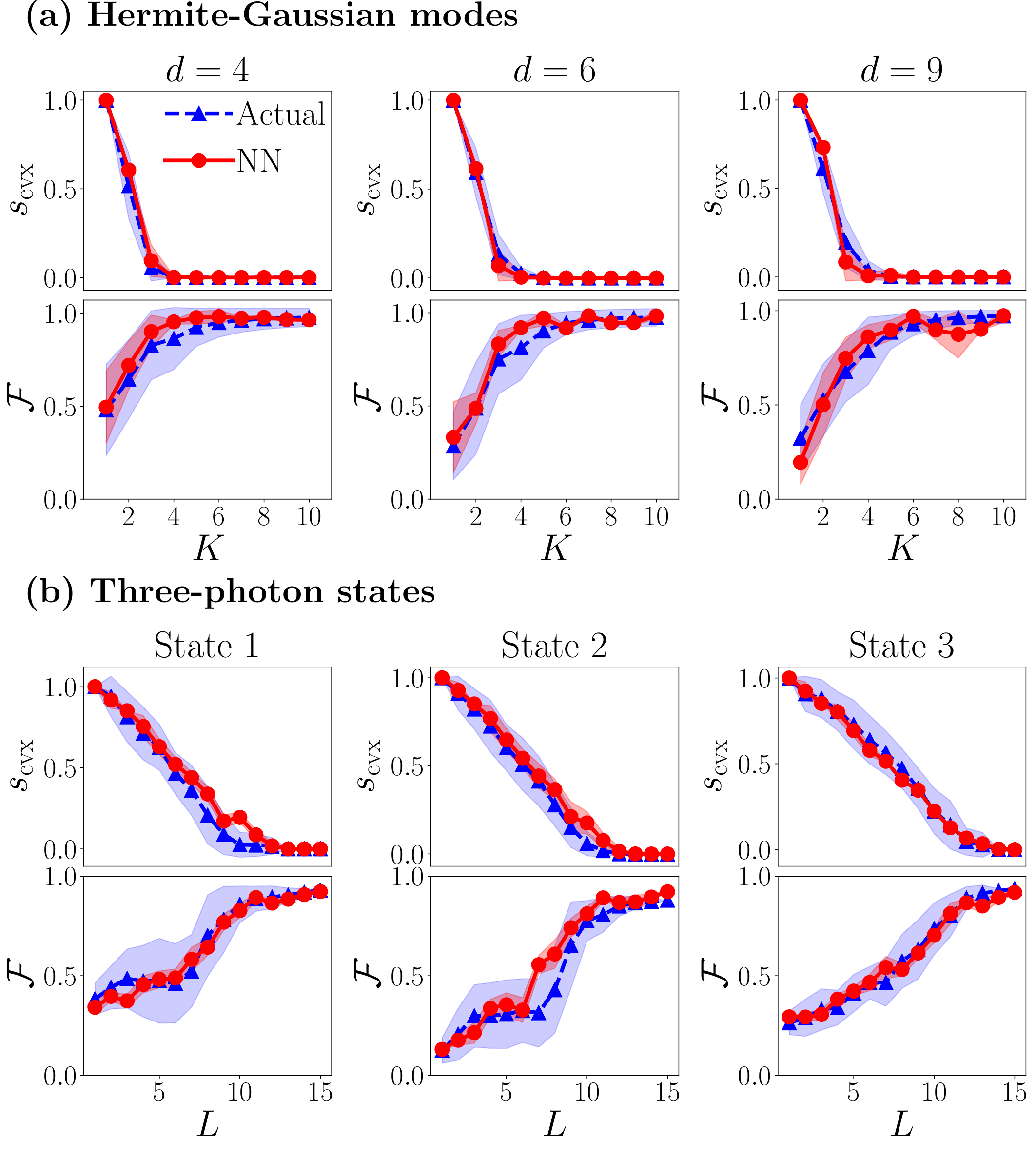}
	\caption{\label{fig:ICCNN_FidNN_expt}(a)~The neural-network predictions of $s_\textsc{cvx}$ and $\mathcal{F}$ for spatial-mode photonic states of dimensions $d=4$, 6 and 9. All graphs and 1-$\sigma$ error bars of each dimension $d$ are obtained from 15 experimental test states used to evaluate the networks. The average fidelity mapped out by FidNet lies closely with the actual computed curve. (b)~Performances on three-photon systems are given for states $\ket{1,2}\bra{1,2}$, $\ket{2,1}\bra{2,1}$, and the rank-two $(\ket{1,2}\bra{1,2}+\ket{2,1}\bra{2,1})/2$ in this order. All graphs and 1-$\sigma$ error bars are obtained from 20 experimental test runs per quantum state. Despite the large error bars of the actual values owing to noise and experimental imperfections, the average fidelity curve is correctly identified by FidNet.}
\end{figure}

\subsection{Experimental performance with spatial-mode photonic states}
\label{subsec:expt1}

For each value of $d$, we experimentally generated random pure states and construct their respective ACT measurement bases in order to evaluate the performance of ICCNet and FidNet, which were previously trained with 10000 simulation datasets of random quantum states of uniformly distributed rank $1\leq r\leq 3$ and different $K$ values. These simulated training datasets are modeled with statistical noise arising from a multinomial distribution defined by $N=5000$ sampling copies per basis, which is close to the experimental average. 

Owing to experimental noise, the resulting spatial-mode quantum states are, as a matter of fact, nearly pure but sufficiently low-rank. Figure~\ref{fig:ICCNN_FidNN_expt}(a) confirms that ICC and fidelity benchmarking with simulation-trained neural network models are accurate even with real experimental test data. One can observe the relative network-prediction stability of $s_\textsc{cvx}$ in contrast with that of $\mathcal{F}$. This coincides with the expectation that while the fidelity is strongly affected by statistical noise and other imperfections such as systematic errors, the degree of informational completeness is more intimately related to the quantum measurements and rank of the quantum state, such that noise only introduces perturbations on the functional behavior of $s_\textsc{cvx}$. Regardless, Fig.~\ref{fig:ICCNN_FidNN_expt}(a) shows that all predictions made by the simulation-trained ICCNet and FidNet models remain roughly within the error margins of actual computed values.

\subsection{Experimental performance with multiphoton states}
\label{subsec:expt2}

For every fixed number ($L$) of projectors chosen from the complete set of 16 defined in Sec.~\ref{subsec:multiphoton}, simulation datasets of 10000 random $d=4$ quantum states of uniformly distributed $r$ are fed into both ICCNet and FidNet for training. These datasets are obtained from randomized sequences of the 16 projectors described above. Statistical noise is introduced into the simulation with multinomial distributions defined by $N=500$ per projective measurement. To test the trained models and acquire prediction results depicted in Fig.~\ref{fig:ICCNN_FidNN_expt}(b), we make use of three different sets of 20 experimental runs outside the training datasets, each set corresponding to a different quantum state. 

\begin{figure}[t]
	\centering\includegraphics[width=1\columnwidth]{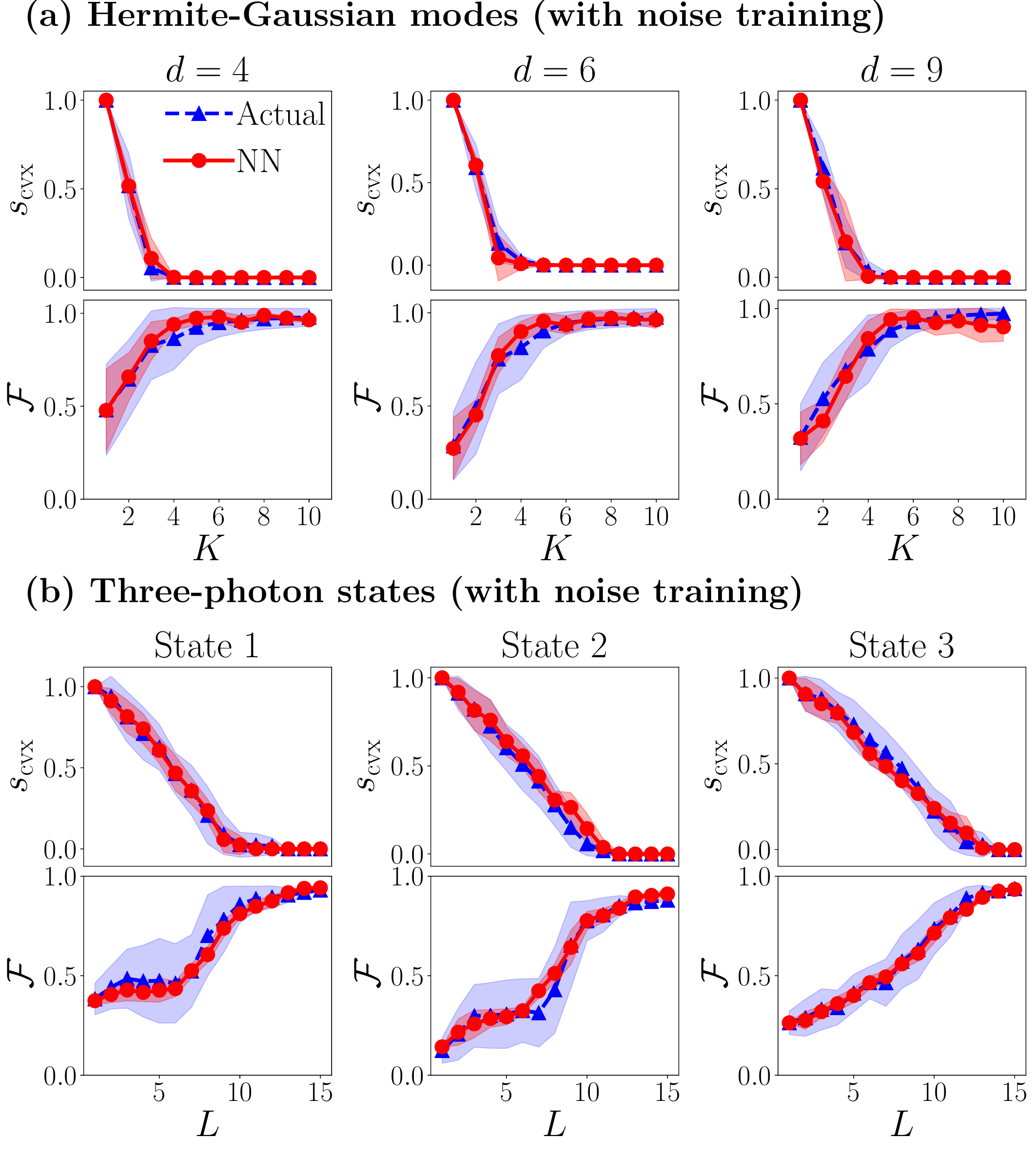}
	\caption{\label{fig:ICCNN_FidNN_expt_boot}The bootstrapped performance of ICCNet and FidNet in predicting $s_\textsc{cvx}$ and $\mathcal{F}$ for the same test datasets that are used in Fig.~\ref{fig:ICCNN_FidNN_expt}, where fluctuating features are generally smoothened with bootstrapped noise training.}
\end{figure}

\subsection{Noise training and reduction}

Experimental noise due to imperfections and systematic errors are always present in any real dataset. Fluctuating deviations of neural-network predicted values from actual ones as observed in Fig.~\ref{fig:ICCNN_FidNN_expt} arise from the lack of such experimental noise in all simulated training datasets, apart from purely statistical fluctuations, used to train ICCNet and FidNet.

When more knowledge about the noisy environment is acquired, data simulation from such knowledge may be carried out to improve the network predictions under such an environment. Here, we show that when some samples of experimental data that are sufficiently representative of the overall noise behavior can be spared for training, it is possible to train ICCNet and FidNet with both statistically-noisy simulated datasets and bootstrapped experimental datasets in order to learn the experimental noise effects approximately well and improve network predictions.

Bootstrapping entails using a given experimental dataset to generate numerous mock datasets using Monte~Carlo procedures. More specifically, in the multinomial setting, the column $\rvec{\nu}_k$ of relative frequencies for the $k$th basis possess a Gaussian distribution of mean $\rvec{p}_k$ and covariance matrix $\dyadic{\Sigma}^{(k)}_{\rvec{p}}=[\mathrm{diag}(\rvec{p}_k)-\rvec{p}_k\,\TP{\rvec{p}_k}]/N$ for sufficiently large $N$ owing to the central limit theorem, where $\mathrm{diag}(\,\cdot\,)$ forms a diagonal matrix whose diagonals are defined by the argument. A direct substitution of $\rvec{\nu}_k$ for $\rvec{p}_k$ leads to the following simple rule for bootstrapping experimental ACT datasets from Hermite-Gaussian mode photonic system:  $\rvec{\nu}'_{k}=\mathcal{N}_{\geq0}\{\rvec{\nu}_{k}+\rvec{w}_{k}\}$, where $\rvec{w}_{k}$ is a column of random variables collectively distributed according to the Gaussian distribution of zero mean and covariance matrix $\dyadic{\Sigma}^{(k)}_{\rvec{\nu}}$, where $\dyadic{\Sigma}^{(k)}_{\rvec{\nu}}$ is to be evaluated with the measurement relative frequencies of the particular $k$th basis and $N$ is set to $5000$, which is the estimated number of copies per ACT basis considered in Sec.~\ref{subsec:expt1}. The operation $\mathcal{N}_{\geq0}$ is a composition of absolute value of the argument followed by its sum normalization over $0\leq j\leq d-1$ for the $k$th ACT basis. Finally, the states that produce the bases relative frequencies used in the bootstrapping procedure are different from the test states used to evaluate the network predictions.

\begin{figure*}[t]
	\centering\includegraphics[width=1.5\columnwidth]{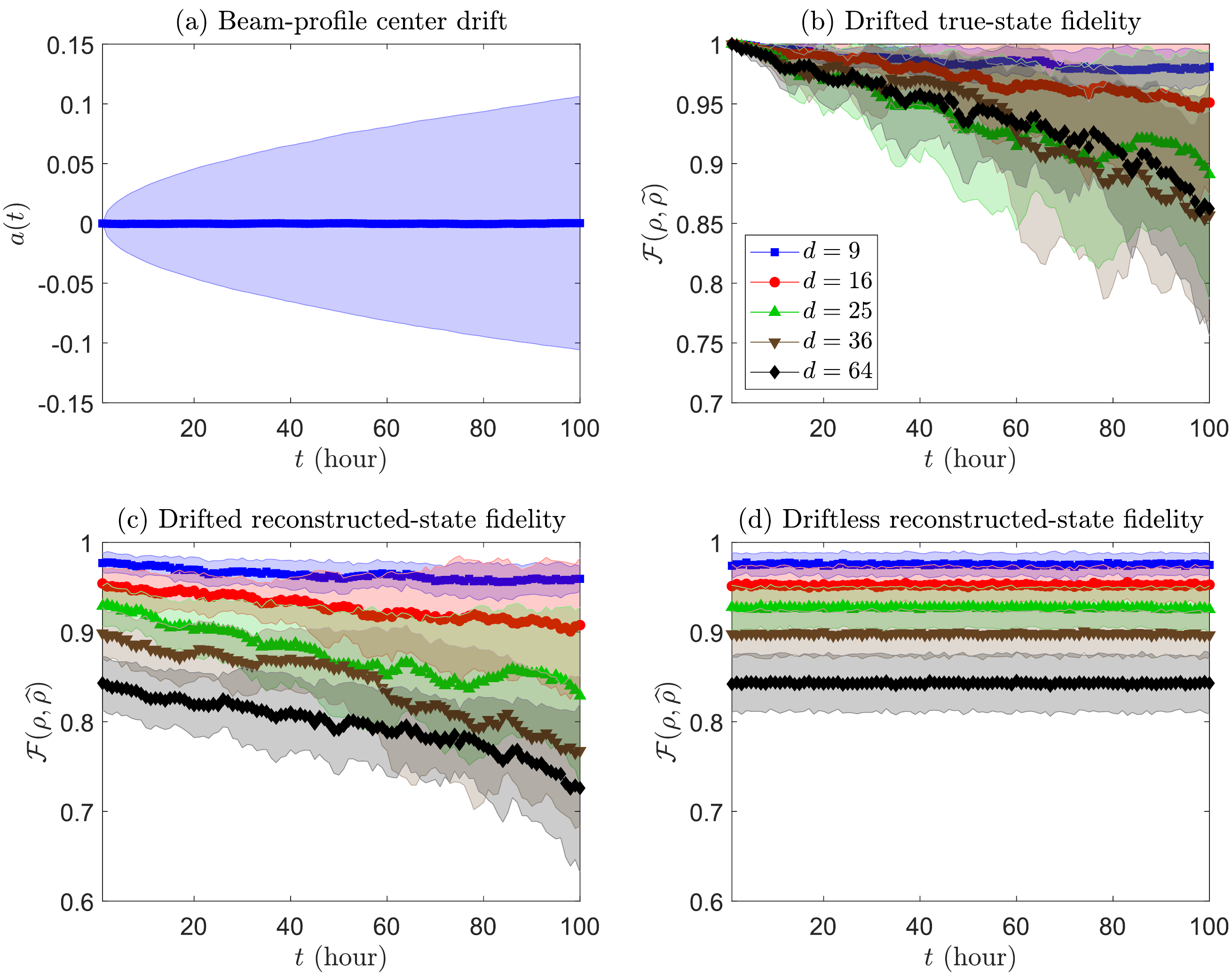}
	\caption{\label{fig:drift_calib}(a)~Random single-direction displacements of the beam-profile center shows an increasing variance around the zero-displacement origin that is expected of random-walk Wiener processes. Such \emph{variance drifts} give rise to the fidelity curves with respect to ideal true states $\rho$ in~(b)~for the noisy true states $\widetilde{\rho}=\widetilde{\rho}(t)$ and those in~(c)~for the reconstructed estimators~$\widehat{\rho}=\widehat{\rho}(t)$, in contrast to the driftless scenarios in (d).}
\end{figure*}

Owing to a limited set of three-photon states, we adopt a different method to bootstrap experimental datasets acquired from these states. Since these datasets are obtained from measuring independent projectors, we randomly permute these projectors and their corresponding relative (unnormalized) frequencies in order generate new measurement sequences as mock datasets. The 16 projectors offer us a total of 16! permutations for each state, allowing us to conveniently generate an abundance of bootstrapped training datasets that are clearly different from those used for testing. By a similar token to the spatial-mode photonic systems, each relative frequency $\nu_{l}$ here is a binomial random variable normalized by the number of copies $N$ used to measure the $l$th projector. Therefore, bootstrapping these relative frequencies may be carried out by additive Gaussian random variables inasmuch as $\nu_{l}'=\nu_{l}+w_{l}\sqrt{\nu_{l}(1-\nu_{l})/N}$, where $w_{l}$ is a standard Gaussian random variable of zero mean and unit variance, and $N=500$ is fixed as the estimated number of copies used to obtain the measured relative frequency for each projector, consistent with Fig.~\ref{fig:ICCNN_FidNN_expt}(b).

Figure~\ref{fig:ICCNN_FidNN_expt_boot} shows the enhanced prediction performances of ICCNet and FidNet. To generate this figure, a total of 5000 simulated and 5000 bootstrapped datasets are employed~($m=10000$) for each group of experiments to train the networks for every value of $K$ and $L$. These new plots indicate that slightly fluctuating neural-network prediction curves on noisy experimental data can be smoothened when bootstrapped information about the noisy environment is incorporated into the training.

\subsection{Suppression of systematic errors}
\label{subsec:drifts}

To end this section, we shall now discuss the implications of all presented results, especially the computation performance graphs shown in Fig.~\ref{fig:time_graphs}, as far as real-time experimental systematic errors are concerned~\cite{Proctor2020,kelly2018physical}. As analytical results are unavailable, we resort to numerical analyses on the effects of ICC-computation-time reduction on such errors. For this purpose, we provide an important example of a kind of systematic drift phenomenon that is highly typical in optical fibers that carry spatial-mode photons, such as those of Hermite-Gaussian modes discussed in this article.

\begin{figure*}[t]
	\centering\includegraphics[width=1.5\columnwidth]{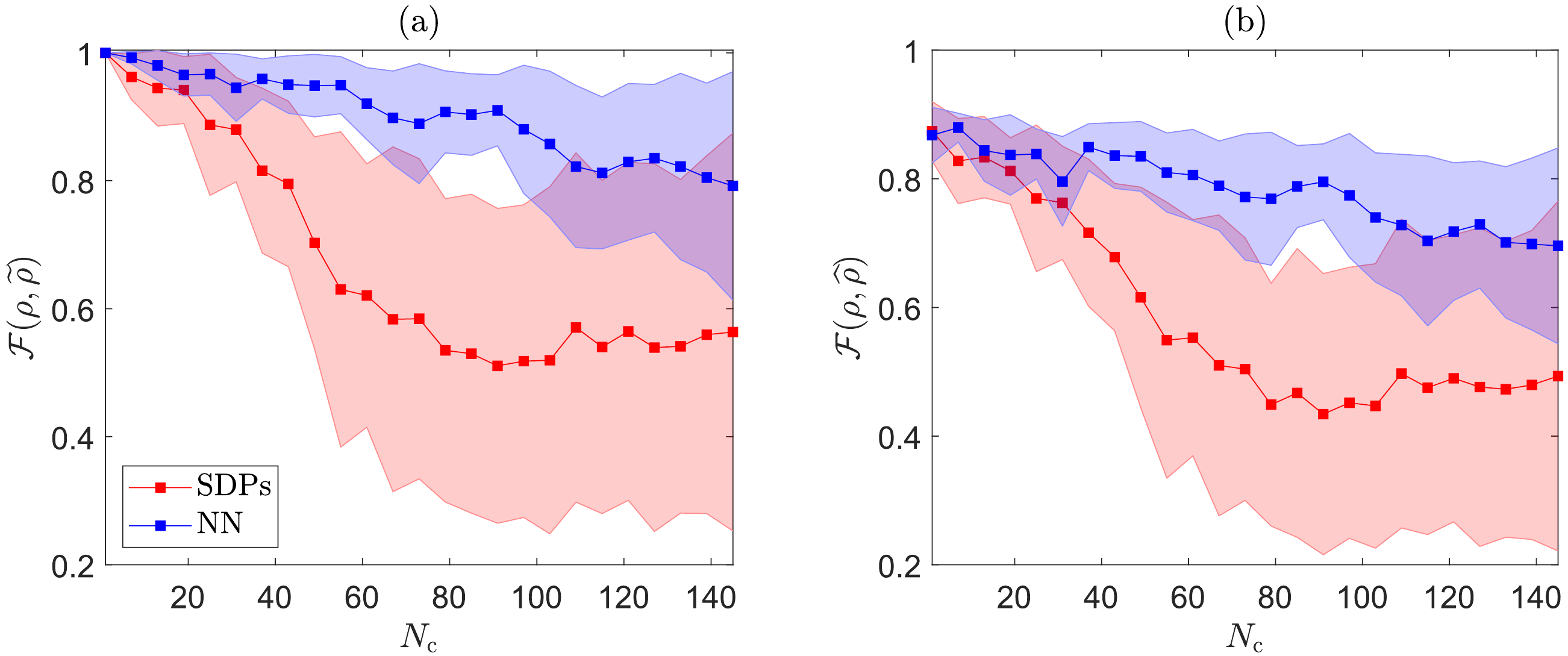}
	\caption{\label{fig:drift_QC}The fidelity $\mathcal{F}$ of the input state $\rho$ with (a)~the noisy true state $\widetilde{\rho}=\widetilde{\rho}(t)$, and that with (b)~its IC estimator $\widehat{\rho}=\widehat{\rho}(t)$, before each round of quantum computation commences (step labeled by $N_\mathrm{c}$) for an eight-qubit $(d=256)$ processor executed with Hermite-Gaussian optical sources. Using ICCNet in place of SDP-based ICC not only \emph{triples} the computation output $N_\mathrm{c}$ in a given period of time (assuming negligible quantum-computation timescales), but also maintains a much more stable fidelity for the same number of computations. All 1-$\sigma$ error regions are computed from 10 different runs.}
\end{figure*}

Focusing only on the transverse plane relative to the propagation direction of a laser beam, a given Hermite-Gaussian mode function $u_m(x)$ of order $m$ in the spatial $x$-coordinate is given by~\cite{Siegman:1986aa}
\begin{equation}
	u_m(x)=\left(\dfrac{2}{\pi}\right)^{1/4}\dfrac{1}{\sqrt{2^m\,m!\,w_0}}\,\HERM{m}{\dfrac{\sqrt{2}x}{w_0}}\,\E{-\frac{x^2}{w_0^2}}\,,
	\label{eq:u_m}
\end{equation}
with $w_0$ being the beam waist, and $\HERM{m}{x}$ a degree-$m$ Hermite polynomial in $x$. Upon using the ket notation, one can express the mode function in the familiar manner inasmuch as $\inner{x}{m;0}_\textsc{hg}=u_m(x)$. An ideal fiber would carry spatial-mode photons of a mode function that has a stable center point, which is usually set at the Cartesian origin as in~\eqref{eq:u_m}. In real experiments, however, a main source of time-dependent systematic errors is transversal displacements of $u_m(x)$ away from the origin~\cite{Hsu_2004,Ndagano:17}, which we may approximately model as random Wiener processes. Such displacements would distort the originally intended true state. Suppose that a basis ket $\ket{m}_\textsc{hg}\equiv\ket{m;0}_\textsc{hg}$ is displaced away from the origin by $a$, it is shown in Appendix~\ref{app:HG_disp} that the resulting displaced $\ket{m;a}_\textsc{hg}=\E{-\I a P}\ket{m;0}_\textsc{hg}$, where $\bra{x}\E{-\I a P}=\bra{x-a}$, possesses the following transformation function  
\begin{align}
	\ket{m;a}_\textsc{hg}=&\,\sum_l\ket{l;0}_\textsc{hg}\,_\textsc{hg}\inner{l;0}{m;a}_\textsc{hg}\,,\nonumber\\
	_\textsc{hg}\inner{l;0}{m;a}_\textsc{hg}=&\,\int\D x'\,u_{m}(x'-a)\,u_{l}(x')\nonumber\\
	=&\,\dfrac{\E{-\frac{a^2}{2w_0}}}{\sqrt{m!\,l!}}\,(-1)^m\,\left(\dfrac{a}{w_0}\right)^{n_>-n_<}\nonumber\\
	&\,\times\KUMMERU{-n_<}{1+n_>-n_<}{a^2/w_0^2}\,,
	\label{eq:HG_transf}
\end{align}
where $n_>=\max\{m,l\}$, $n_<=\min\{m,l\}$, and
\begin{align}
	&\,\KUMMERU{-n_<}{1+n_>-n_<}{x^{-2}}\nonumber\\
	=&\,x^{-2n_<}\sum^{n_<}_{l=0}\,(-x^2)^l\,\binom{n_<}{l}\,\dfrac{n_>!}{(n_>-l)!}
\end{align}
is Kummer's confluent hypergeometric function. This result reduces to that in~\cite{Hsu_2004} for the $m=0$ special case. The complete two-dimensional transverse profile of a Hermite-Gaussian mode in space is therefore described by the mode function $u_{m,n}(x,y)\equiv u_m(x)u_n(y)$, where the corresponding displaced kets $\ket{m,n;\rvec{a}}_\textsc{hg}\equiv\ket{m;a_1}_\textsc{hg}\ket{n;a_2}_\textsc{hg}$ exhibit the transformation
\begin{equation} \ket{m,n;\rvec{a}}_\textsc{hg}=\sum_{l,l'}\ket{l,l';\rvec{0}}_\textsc{hg}\,_\textsc{hg}\inner{l;0}{m;a_1}_\textsc{hg}\,_\textsc{hg}\inner{l';0}{n;a_2}_\textsc{hg}\,,
\end{equation}
where the coefficients are pair-products of one-dimensional transformation functions as in \eqref{eq:HG_transf}. Components of the two-dimensional displacement $\rvec{a}=(a_1\,\,\,a_2)^\top$ are assumed to be independent. After a period of time $t$, the noisy true state $\widetilde{\rho}=\sum_{m,n,m',n'}\ket{m,n;\rvec{a}(t)}_\textsc{hg}\,\rho_{mn,m'n'}\,_\textsc{hg}\bra{m',n';\rvec{a}(t)}$ is now defined in the displaced Hermite-Gaussian basis of some displacement $\rvec{a}(t)$. The same type of disturbances apply to the POVM outcomes since they are implemented digitally using SLMs with the same beams.

In the experiments that gathered data used in plotting Fig.~\ref{fig:ICCNN_FidNN_expt}(a), the root-mean-square displacement of the Hermite-Gaussian beam center was measured to be about 5\% of the beam waist $w_0$ after a period of 24 hours. This approximately coincides with a model of a Wiener process specified by the random displacement variable $a(t)=a(t-1)+b(t-1)$, which is a cumulative temporal sum of random variables $b(t)$ that are each distributed according to the standard Gaussian distribution defined by the standard deviation $\sigma\approx w_0/95$ when the time coordinate $t$ is in units of an hour. Figure~\ref{fig:drift_calib} shows the Hermite-Gaussian beam-profile center displacement and fidelity curves in time $t$ for various dimensions $d$, where $d=d_0^2$ is the product of the individual dimensions $d_0$ of the truncated Hilbert space spanned by the finite set of Hermite-Gaussian basis kets of orders $0\leq m,n\leq d_0-1$. In actual experiments, there most likely exist other time-dependent sources of errors not modeled here that could worsen the fidelities.

To appreciate the significance of systematic-error suppression, we consider a quantum processor that utilizes Hermite-Gaussian beams as a source for producing high-dimensional initial states for quantum computation. Suppose that the processor is running continuously under server conditions, and maintenance is carried out before significant systematic drifts are anticipated. Whenever the processor refreshes after each set of computations is completed, the newly prepared initial state $\rho$ undergoes compressive state tomography to ensure that it is within the expected error margins.

As an instructive example, we consider at an eight-qubit processor controlled by a $d=256$ Hermite-Gaussian source~\cite{manip:2012}. Each projector exposure time is about 1.5~seconds~(sec), so that the measurement time of $K$ von~Neumann bases is $t_\mathrm{meas}=1.5\,Kd=1920$~sec. Preliminary calibration indicates that such an exposure period yields $N=1000d=2.56\times10^5$ state copies per basis. For a realistic error modeling, the expressions of $t_\mathrm{meas}$ and $N$ in $d$ were calibrated from the actual setup used to collect our experimental data for Figs.~\ref{fig:ICCNN_FidNN_expt} and \ref{fig:ICCNN_FidNN_expt_boot}. We focus on the preparation of pure states, each of which requires only $K=5$ random bases for an IC reconstruction~\cite{Ahn:2019ns}, the average time, estimated over five random sets of $K=5$ bases, for an ICC verification~(two SDPs) is $t_\textsc{sdp}=4000$~sec using the personal computer with hardware specification given in the caption of Fig.~\ref{fig:time_graphs}. After carrying out the basis measurements and ICC, the final state estimator $\widehat{\rho}$ is given by the ML estimator that takes an average of $t_\textsc{ml}\approx 240$~sec to generate using the accelerated projected-gradient algorithm in~\cite{Shang:2017sf}, which is insignificant in comparison to $t_\textsc{sdp}$. The time for each round of quantum computation depends on the actual application. For simplicity, we assume that each round of quantum computation is executed almost instantly since no classical post-processing is needed. From these specifications, we note that $t_\textsc{sdp}\approx 2 t_\mathrm{meas}$, and the prefactor grows with $d>256$. Therefore, in the case of $d=256$, replacing the two SDP algorithms in ICC with a trained ICCNet would shave about 66\% of the total computation time off. For larger dimensions, if so desired, the ML estimation procedure may be completely replaced with trained conditional generative networks~\cite{ahmed2020quantum,ahmed2020classification} to eliminate $t_\textsc{ml}$.

Figure~\ref{fig:drift_QC} shows the fidelity of the estimator $\widehat{\rho}$ with the generated initial state $\rho$ before every quantum-computation step ($N_\mathrm{c}$ of them in total). The graphs highlight the adverse effects of systematic errors when initial-state verification takes too long. Hence, compressive tomography performed with trained neural networks provides a better solution to real-time device certification with a higher fidelity stability, so that quantum computation can run much more smoothly with a greater $N_\mathrm{c}$ output before drift maintenance is applied.

\section{Concluding Remarks}

We took advantage of the universality of convolutional networks to train two neural networks that can very efficiently certify a low-measurement-cost quantum-state characterization scheme. These networks can respectively benchmark the quantum completeness of a given set of measurement outcomes and corresponding data for reconstructing an unknown quantum state, as well as the resulting fidelity without explicitly carrying out the state reconstruction. Our machine-learning-assisted scheme therefore allows experimentalists to rapidly assess the sufficiency of measurement resources for an unambiguous characterization of arbitrary quantum states and achieve accelerated real-time verification without having to perform any optimization routine during the experiment. This becomes essential for many practical quantum tasks that do require fast execution times to avoid noise accumulation and drifts.

An arguably interesting problem would be to minimize the time required to acquire the trained neural networks. This includes both the training time and the generation of adequate training datasets. While the former can now be easily parallelized with graphics processing units, the latter involves two rounds of semidefinite programming per dataset as discussed in Sec.~\ref{subsec:aIC}, the acceleration of which is still a subject of ongoing research~\cite{Majumdar:2020aa}. 

On the other hand, while classical algorithms for these procedures have worst-case polynomial time complexities in the dimension of the Hilbert space, it is known~\cite{brando_et_al:LIPIcs:2019:10603} that quantum algorithms can execute semidefinite programs with polylogarithmic time complexities in the dimension. This immediately reveals the possibility of completely transforming the neural networks employed here, or part thereof, into their quantum counterparts (fused with the training-data processing procedures that use quantum semidefinite programming) that could assimilate into a much larger set of networks for a grander purpose. Practical feasibility in implementing such extended quantum neural networks still remains to be seen.

\emph{Note.---}Nearing the submission of our work, we discovered another very recent preprint reference~\cite{zhang2021direct} that purely discusses the estimation of the fidelity with fully-connected networks. Apart from the clear distinction in architectures, we remark that while the networks in this reference were specifically trained for Pauli measurements, our CNN-based FidNet is compatible with generalized measurement inputs that can be readily used in parallel with ICCNet or any other quantum task that relies on arbitrary measurements. The objectives of both works are hence very different. The next key distinction is network training, which in this reference is based on categorical training that splits the continuous fidelity range into small intervals. As mentioned in the preprint itself, network training can be slow when the intervals are too small. In our current work, FidNet directly computes the fidelity values without such output splitting, and hence training efficiency is not sacrificed for prediction accuracy.

\begin{acknowledgments}
	Y.S.T., S.S. and H.J. acknowledge support by the National Research Foundation of Korea (Grant Nos. 2019R1A6A1A10073437, 2019M3E4A1080074, 2020R1A2C1008609, and 2020K2A9A1A06102946) via the Institute of Applied Physics at Seoul National University, and by the Institute of Information \& Communications Technology Planning \& Evaluation (IITP) grant funded by the Korea government~(MSIT)~(Grant Nos. 2020-0-01606 and 2021-0-01059). G.L. acknowledges support by the  Center of Excellence $\ll$Center of Photonics$\gg$ funded by the Ministry of Science and Higher Education of the Russian Federation, contract No. 075-15-2020-906. Y.K. and Y.-H.K. acknowledge support by the National Research Foundation of Korea (Grant No. 2019R1A2C3004812) and the ITRC support program (IITP-2020-0-01606). L.L.S.S. acknowledges support from European Union's Horizon 2020 research and innovation program (ApresSF and STORMYTUNE) and the Ministerio de Ciencia e Innovaci{\'o}n (PGC2018-099183-B-I00). The MSU team acknowledges support from the Russian Foundation for Basic Research (RFBR Project No. 19-32-80043 and RFBR Project No. 19-52-80034) and support under the Russian National Technological Initiative via MSU Quantum Technology Centre. SSS and SPK acknowledge	support by the Development Program of the Interdisciplinary Scientific and Educational School of Lomonosov Moscow State University `Photonic and quantum technologies: Digital medicine'.
\end{acknowledgments}

\appendix

\vspace{7ex}
\hrule
\vspace{1ex}
\noindent
\emph{* All tables and figures for the appendix are found behind the bibliography section.}

\section{Procedures for generating random measurements}
\label{app:gen_rand_meas}

We state the recipes for generating two types of measurement bases used to generate all training datasets discussed in Sec.~IV. We start with the set of von~Neumann bases derived from rotations with random unitary operators $U^\mathrm{Haar}_k$ distributed according to the Haar measure. Such a set of unitary operators was shown to be derivable from a modified version of the QR~decomposition~\cite{Mezzadri:2007qr}:

\begin{center}
	\begin{minipage}[c][7cm][c]{0.9\columnwidth}
		\noindent
		\rule{\columnwidth}{1.5pt}\\
		\textbf{Constructing a random Haar basis}\\[1ex]
		Starting from a reference basis $\{\ket{0},\ket{1},\ldots,\ket{d-1}\}$:
		\begin{enumerate}
			\item Generate a random $d\times d$ matrix $\dyadic{A}$ with entries i.i.d. standard Gaussian distribution.
			\item Compute the two matrices $\dyadic{Q}$ and $\dyadic{R}$ by searching for the QR decomposition $\dyadic{A}=\dyadic{Q}\dyadic{R}$.
			\item Define $\dyadic{R}_\text{diag}=\mathrm{Diag}\{\dyadic{R}\}$ (sets all off-diagonal elements to zero) 
			\item Define $\dyadic{L}=\dyadic{R}_\text{diag} \oslash |\dyadic{R}_\text{diag}|$ ($\oslash$ refers to the Hadamard division).
			\item Define the new basis $\ket{u_l}\widehat{=}\,\dyadic{U}_\mathrm{Haar}\ket{l}$ for $0\leq l\leq d-1$.\\[-5ex]
		\end{enumerate}
		\rule{\columnwidth}{1.5pt}
	\end{minipage}
\end{center}

The next kind of measurement is an adaptive set of von~Neumann bases that are sequentially inferred directly from previous measurement data. They were meant for establishing an ACT scheme that is highly compressive~\cite{Ahn:2019aa,Ahn:2019ns}. The logic behind their construction is that low-rank true states are relatively closer to rank-deficient state estimators, and minimum-entropy estimators obtained from non-IC bases set give a very compressive sequence of eigenbases to quickly reach informational completeness. We state the construction of such a set of $K$ bases $\{\mathcal{B}_1,\mathcal{B}_2,\ldots,\mathcal{B}_K\}$ below, with the first basis $\mathcal{B}_1=\{\ket{l}\bra{l}\}_{l=0}^{d-1}$ being the standard computational basis:

\begin{center}
	\begin{minipage}[c][10.5cm][c]{0.9\columnwidth}
		\noindent
		\rule{\columnwidth}{1.5pt}\\
		\textbf{Constructing an ACT basis}\\
		\noindent
		Beginning with $k=1$ and a random computational basis $\mathcal{B}_1$:
		\begin{enumerate}
			\item Measure $\mathcal{B}_k$ and collect the relative frequency data $\sum^{d-1}_{j'=0}\nu_{j'k}=1$.
			\item From $\left\{\nu_{0k'},\ldots,\nu_{d-1\,\,k'}\right\}^k_{k'=1}$, obtain $kd$ physical probabilities.
			\item Perform ICC with the physical probabilities and compute $s_{\textsc{cvx},k}$:
			\begin{itemize}
				\item \textbf{If}~$s_{\textsc{cvx},k}<\epsilon$, terminate this ACT scheme and take $\rho_\text{max}\approx\rho_\text{min}$ as the estimator and report $s_{\textsc{cvx},k}$.
				\item \textbf{Else}~Proceed.
			\end{itemize}
			\item Choose an estimator $\widehat{\rho}_k$ that minimizes the von~Neumann entropy $S(\widehat{\rho}_k)=-\tr{\widehat{\rho}_k\log \widehat{\rho}_k}$ subject to the positivity and data constraints.
			\item Define $\mathcal{B}_{k+1}$ to be the eigenbasis of $\widehat{\rho}_k$.
			\item Set $k=k+1$ and repeat.\\[-5ex]
		\end{enumerate}
		\rule{\columnwidth}{1.5pt}
	\end{minipage}
\end{center}

\section{Hyperparameters of ICCNet and FidNet}
\label{app:hyper}

All hyperparameters used in both ICCNet and FidNet are manually optimized so that the validation loss is minimized (see Appendix~\ref{app:explicit}). Figure~\ref{fig:hyperparam_ICCNet} succinctly consolidates all important operational hyperparameter settings adopted for training ICCNet in various dimensions. Networks~(a) and (b) have identical architecture with different dropout rates. For larger dimensions such as $d=64$, we find that the use of deeper CNN networks can yield better training results. In these cases, \emph{residual blocks} implemented in network~(c) help to avoid the so-called vanishing-gradient problem~\cite{He:2016aa}. Each residual block consists of repeated convolutional blocks ({\bf BLK}$_\mathrm{s}(t)$) that maintain the array dimensions, sandwiched by a skip connection that adds the input of these repeated convolutional blocks to their output. Another notable difference between Fig.~\ref{fig:hyperparam_ICCNet}(c) and Figs.~\ref{fig:hyperparam_ICCNet}(a) and (b) is the inclusion of an average-pooling layer and one fully-connected layer, which are standard components of residual-based networks. Figure~\ref{fig:hyperparam_FidNet} shows the corresponding hyperparameters for FidNet.

\section{Explicit network training procedures}
\label{app:explicit}

\subsection{Input data preparation}
\label{subsec:input_prep}

The initial input data matrix $\dyadic{X}$ for training the ICCNet comprises the $m$ datasets of $K$ measurement bases, or $L$ projectors, and their corresponding relative frequencies. These data are reshaped into either a $m\times\lceil \sqrt{K(d^2+d)}\rceil \times \lceil \sqrt{K(d^2+d)}\rceil$ or $m\times\lceil\sqrt{L(d^2+1)}\rceil\times\lceil\sqrt{L(d^2+1)}\rceil$ three-dimensional matrix $\widetilde{\dyadic{X}}$ to be processed by the convolution networks. 

The input data matrix for training the FidNet requires the additional $m$ target states to be assigned to the respective datasets. For $d=16$, 32 and 64, FidNet is trained by supplying the true states as the (``right'') target states. This is sufficient as only statistical fluctuation exist in the simulation data obtained from finite copies. To benchmark fidelities for real experimental data, it is important that FidNet also recognizes inputs with systematic errors. We numerically show that training FidNet with both the ``right'' and ``wrong'' target states, the latter referring to targets differing from the true states for the same datasets, can improve the benchmarking accuracy on average. Combining these datasets give the resulting reshaped three-dimensional matrix that is either of size $2m\times\lceil \sqrt{(K+1)d^2+Kd}\rceil \times \lceil \sqrt{(K+1)d^2+Kd}\rceil$ or $2m\times\lceil\sqrt{(L+1)d^2+L}\rceil\times\lceil\sqrt{(L+1)d^2+L}\rceil$. As a demonstration, we take ``wrong'' target state $\rho_\text{wrong}$ to be a randomly generated operator from the true state $\rho\,\widehat{=}\,\dyadic{U}\,\dyadic{\Lambda}\,\dyadic{U}^\dag$ diagonalized with the unitary matrix $\dyadic{U}$ and diagonal matrix $\dyadic{\Lambda}$ according to the following prescription:\\[1ex]

For a $d$-dimensional column $\rvec{v}$ of uniformly-distributed entries, each in the range [0,1], define $\rho_\text{wrong}\,\widehat{=}\,\dyadic{U}'\dyadic{\Lambda}'\dyadic{U}'^\dag$ using the matrices
\begin{align}
\dyadic{\Lambda}'=&\,\mathrm{Diag}\!\left((1-\lambda_1)\,\mathrm{diag}(\dyadic{\Lambda})+\lambda_1\rvec{w}\right)\,,\nonumber\\
\dyadic{U}'=&\,\dyadic{V}\dyadic{U}\,,\nonumber\\
\rvec{w}=&\,\mathcal{N}\left\{-\log(\rvec{v})\odot\mathrm{diag}(\dyadic{\Lambda})\right\}\,,\nonumber\\
\dyadic{V}=&\,\mathrm{wHaar}(\lambda_2)\,,
\end{align}
where $\lambda_1$ and $\lambda_2$ are uniformly distributed in [$\lambda_\text{min}$,$\lambda_\text{max}$], $\mathrm{diag}(\,\bm{\cdot}\,)$ and $\mathrm{Diag}(\,\bm{\cdot}\,)$ are respectively diagonal-element extracting and diagonal-matrix transforming operations, $\mathcal{N}\{\,\bm{\cdot}\,\}$ normalizes a column by its element-wise sum, $\odot$ denotes the Hadamard product, and $\mathrm{wHaar}(\lambda)$ refers to the weighted Haar unitary function that outputs a random unitary according to the assigned weight $\lambda$. By definition, the special case $\mathrm{wHaar}(0)=\dyadic{1}$ holds, and the more general function is given by
\begin{center}
	\begin{minipage}[c][6cm][c]{0.9\columnwidth}
		\noindent
		\rule{\columnwidth}{1.5pt}\\
		\textbf{Weighted Haar unitary (wHaar) of weight $\lambda$}
		\begin{enumerate}
			\item Generate a random $d\times d$ matrix $\dyadic{A}$ with entries i.i.d. standard Gaussian distribution.
			\item Define $\dyadic{A}'=\lambda\,\dyadic{A}+(1-\lambda)\,\dyadic{1}$.
			\item Compute $\dyadic{Q}$ and $\dyadic{R}$ from the QR decomposition $\dyadic{A}=\dyadic{Q}\dyadic{R}$.
			\item Define $\dyadic{R}_\text{diag}=\mathrm{Diag}\{\dyadic{R}\}$ and $\dyadic{L}=\dyadic{R}_\text{diag} \oslash |\dyadic{R}_\text{diag}|$ ($\oslash$ refers to the Hadamard division).
			\item Define $\dyadic{V}=\dyadic{Q}\dyadic{L}$.\\[-5ex]
		\end{enumerate}
		\rule{\columnwidth}{1.5pt}
	\end{minipage}
\end{center} 
\noindent
It is clear that $\rho_\text{wrong}$ so defined has exactly the same rank as $\rho$, and at times can be close to $\rho$. A practical justification for these sort of target states is that very typically in experiments, although the target states are not exactly $\rho$ due to various noisy imperfections, the actual true states are, nevertheless, very often almost as rank-deficient as the intended target states, with a rapidly decaying eigenvalue spectrum. Figures~\ref{fig:FidNet_compare} and \ref{fig:FidNet_compare2} show that fidelity benchmarking is typically optimal when training is performed with both the ``right'' and ``wrong'' target states simultaneously. For these experimental data, we find that $\lambda_1=0.8$ and $\lambda_2=1$ gives rather accurate fidelity benchmarking. For more general noisy situations, we may need to introduce more structured noise models in generating the simulated training datasets.

\subsection{Training validation}
\label{subsec:val}

The specifications of every input data matrix are tabulated in Tab.~\ref{tab:data}. Generally speaking, the action of training these convolutional networks is equivalent to carrying out an optimization routine to minimize the ``distance'', quantified by a so-called loss function, between the predicted output~$\rvec{y}_\mathrm{pred}$ and the original training output~$\rvec{y}$. In all training procedures, the momentum-based gradient-descent algorithm NAdam~\cite{Dozat:2016aa} is employed for the minimization. The batch size is chosen to strike a compromise between training convergence and gradient-computation accuracy. For ICCNet, we consider the mean absolute-error~(MAE) as the loss function for training. For FidNet, the mean squared-error~(MSE) loss is used.

It is important to track the training activities so that overtraining or overfitting does not happen. To do this, a very common way is to first split the complete data ($\widetilde{\dyadic{X}}$,$\rvec{y}$) into data for training ($\widetilde{\dyadic{X}}_\mathrm{train}$,$\rvec{y}_\mathrm{train}$), validation ($\widetilde{\dyadic{X}}_\mathrm{val}$,$\rvec{y}_\mathrm{val}$) and testing ($\widetilde{\dyadic{X}}_\mathrm{test}$,$\rvec{y}_\mathrm{test}$). During training, at each epoch~(gradient-descent iterative step), as the loss function between $\rvec{y}_\mathrm{pred}$ and $\rvec{y}_\mathrm{train}$ is reduced, the resulting validation loss, that is loss between $\rvec{y}_\mathrm{pred}$ and $\rvec{y}_\mathrm{val}$, is also reported. Training is successful when both training and validation losses decay simultaneously with the number of epochs. As an additional precaution, we confirm both the training and validation progress by performing one final prediction with $\rvec{y}_\mathrm{test}$ to verifying that the test loss is also comparatively small. In our context, the split ratio between training, validation and test datasets is set to 0.8:0.1:0.1. 

For all the training datasets reflected in Tab.~\ref{tab:data}, each row of $\dyadic{X}$ consists of information about the POVM and relative frequencies for the case of ICCNet (and an additional target state for FidNet) that originate from a randomly generated quantum state of rank $r\in[1,3]$. The distribution of continuous entries in the output $\rvec{y}$ can also affect training efficiency. In the case of $s_\textsc{cvx}$ output for ICCNet, there can coexist two groups of values, one group containing values that are substantially far away from zero and the other containing those that are almost zero~(IC). We find that a reversible mapping that maps each output value $y\equiv s_\textsc{cvx}$ to $y'=-\log_{10}(y)/10$, with the conditional definition $y<10^{-10}\rightarrow y=10^{-10}$ to ensure that $y'\leq1$, can improve training efficiencies. All ICCNet architectures shown in Fig.~\ref{fig:hyperparam_ICCNet} and ICCNet training graphs in Fig.~\ref{fig:train_val} refer to these logarithmized outputs. One can understand this logarithmic training as a switch of training focus to order of magnitude estimation for $s_\textsc{cvx}$, which is an alternatively relevant outcome since all one really needs to know is whether a given measurement is IC or not. 

\subsection{Training results analyses}

After every network training, only the model weights corresponding to the lowest validation loss are saved for later predictions. As sample illustrations, we explicitly show the progress of training and validation losses for $d=16$ in Fig.~\ref{fig:train_val}. For this dimension, the input datasets of all four data types listed in Tab.~\ref{tab:data} are stacked for training ICCNet and FidNet at one go. In order to verify the test accuracies offered by the trained neural-network models, we also supply Figs.~\ref{fig:scvx_test} and \ref{fig:fid_test} for $K=4$ assorted von Neumann bases. For completeness, we also furnish numerical performance indicators for ICCNet and FidNet in Tabs.~\ref{tab:iccnet} and \ref{tab:fidnet} to supplement Figs.~7, 9 and 10 in the main text.

As far as the analyses of the training results are concerned, the aforementioned figures and tables are sufficient to verify the training qualities of ICCNet and FidNet. Going by a different route, one may additionally fall back on other more conventional tools to analyze the neural-network-predicted $s_\textsc{cvx}$ values. In theory, the measurements are IC when $s_\textsc{cvx}$ is zero. In practice, however, we assign a small threshold that distinguishes datasets that are IC from those that are not. In Fig.~\ref{fig:CM}, we present the so-called \emph{confusion matrix} that simplistically quantifies how well the datasets are correctly grouped into the ``IC'' and ``non-IC'' classes. The threshold is set at $10^{-3}$. Clearly, the ideal prediction result is such that the off-diagonal elements of the confusion matrix are all zero. Such a perfect binary classification does not exist in realistic machine learning applications. Instead, upon labeling the IC cases as the ``positives'', there would be predictions that are false positives~(fp) (as opposed to the true positives~tp) or false negatives~(fn) (as opposed to the true negatives~tn), giving rise to nonzero, but small, off-diagonal values. We define the concepts of \emph{precision} $\text{prec}=\text{tp}/(\text{tp}+\text{fp})$ and \emph{recall} $\text{rec}=\text{tp}/(\text{tp}+\text{fn})$, and a predictive ICCNet should generally output values that have high precision and recall. More specifically, there exists the so-called F1 score, or more appropriately the S{\o}rensen–Dice coefficient~\cite{Sorensen:1948aa,Dice:1945aa}, $F_1=2\,\text{tp}/(2\,\text{tp}+\text{fp}+\text{fn})$ defined as the harmonic mean of $\text{pr}$ and $\text{rec}$ that speculates such a binary prediction power.

Rather than fixing a particular threshold, it is more objective to scan a range of threshold values and parametrically plot the so-called precision-recall~(PR) curves as shown in Fig.~\ref{fig:PR-ROC}. Then a natural figure of merit to gauge the binary classification power would be the ``area-under-curve''~(AUC) measure for these curves, since a unit area entails the largest possible coverage of prec and rec. This figure of merit also possesses one crucial advantage. If we remember that all training datasets are obtained from random states of uniformly distributed ranks $r\in[1,3]$, it is then easy to see that the $K=4$ datasets, for instance, have much fewer positive cases as compared to negative cases. Such a class imbalance biases the binary classification analysis, and occurs ubiquitously in our context since informational completeness is rank-sensitive, and thus highly dependent on the state ranks used to generate the training datasets. The AUC for the PR curve is consequently lowered mainly because of this bias rather than a weak binary-classification capability. In such cases, perhaps a better option would be to investigate the AUC of the so-called receiver operating characteristic~(ROC) curve~\cite{Branco:2016aa}. This plots the true positive rate~($\text{tpr}=\text{tp}/\text{total positives}$) against the false positive rate~($\text{fpr}=\text{fp}/\text{total negatives}$). For such imbalanced cases, while the PR curve takes a larger $K$ value to recuperate its area, the AUC of the ROC curve remains high for all tested $K$ values~(see Fig.~\ref{fig:PR-ROC}). A loosely, yet intuitive understanding is that the ROC curve accounts for both tp and fp values evenly, whereas the PR curve focuses only the tp values, which form the minority class in a class imbalance situation.

Despite the above observations, just like any other single-number criterion that is popularly adopted in statistics and machine learning owing to its computation simplicity, these figures of merit are \emph{ad hoc} by nature. Therefore, care must be taken in interpreting these measures. 

\section{Mode transformation function of a displaced Hermite-Gaussian mode of arbitrary order}
\label{app:HG_disp}

We proceed to calculate the integral
\begin{equation}
	I_{m,n,a}=\int\D x'\,u_{m}(x'-a)\,u_{n}(x')\,,
	\label{eq:I_m_n_a}
\end{equation}
where
\begin{equation}
	u_m(x)=\left(\dfrac{2}{\pi}\right)^{1/4}\dfrac{1}{\sqrt{2^m\,m!\,w_0}}\,\HERM{m}{\dfrac{\sqrt{2}x}{w_0}}\,\E{-\frac{x^2}{w_0^2}}\,.
\end{equation}
It turns out that the closed-loop integral representation 
\begin{equation}
	\HERM{n}{y}=\dfrac{n!}{2\pi\I}\,\oint_\mathbf{C}\D z\,\dfrac{\E{2yz-z^2}}{z^{n+1}}
\end{equation}
of a degree-$n$ Hermite polynomial in the complex plane is extremely useful for this endeavor, where the contour $\mathbf{C}$ is a closed loop encircling the origin. This yields the triple integrals
\begin{align}
	I_{m,n,a}=&\,\sqrt{\dfrac{2}{\pi w_0^2}}\dfrac{1}{\sqrt{2^{m+n}\,m!\,n!}}\dfrac{m!\,n!}{(2\pi\I)^2}\nonumber\\
	&\,\times\,\oint_\mathbf{C}\D z\,\dfrac{\E{-z^2}}{z^{m+1}}\,\oint_\mathbf{C}\D z'\,\dfrac{\E{-z'^2}}{z'^{n+1}}\,I'_{z,z',a}\,,
\end{align}
which involves another simple Gaussian integral
\begin{align}
	I'_{z,z',a}=&\,\E{-\frac{a^2}{w_0^2}-\frac{2\sqrt{2}z}{w_0}\,a}\!\!\int\!\D x'\,\E{-\frac{2}{w_0^2}x^2+\left[\frac{2a}{w_0^2}+\frac{2\sqrt{2}}{w_0}\left(z+z'\right)\right]\!x}\,,\nonumber\\
	=&\,\sqrt{\dfrac{\pi w_0^2}{2}}\,\E{-\frac{a^2}{w_0^2}-\frac{2\sqrt{2}a}{w_0}\,z}\,\E{\frac{w_0^2}{8}\left[\frac{2a}{w_0^2}+\frac{2\sqrt{2}}{w_0}(z+z')\right]^2}\,.
\end{align}

We implicitly consider the case where $m\leq n$ and first evaluate the $z'$ integral of $I_{m,n,a}$. After some simplification of exponential functions, an application of Cauchy's residue theorem immediately gives
\begin{widetext}
\begin{align}
	&\,\dfrac{n!}{2\pi\I}\,\oint\D z'\dfrac{\E{-z'^2}}{z'^{n+1}}\,I'_{z,z'a}=\sqrt{\dfrac{\pi w_0^2}{2}}\,\E{-\frac{a^2}{2w_0^2}-\frac{\sqrt{2}a}{w_0}\,z+z^2}\!\!\left.\left(\dfrac{\partial}{\partial z'}\right)^{\!\! n}\!\!\E{\left(\frac{\sqrt{2}a}{w_0}+2z\right)z'}\right|_{z'=0}\!\!\!\!\!=\sqrt{\dfrac{\pi w_0^2}{2}}\,\E{-\frac{a^2}{2w_0^2}-\frac{\sqrt{2}a}{w_0}\,z+z^2}\left(\frac{\sqrt{2}a}{w_0}+2z\right)^{\!\! n}\,.
\end{align}
\end{widetext}
We are now ready to finalize the calculation for $m\leq n$ by evaluating the remaining $z$ integral, and a second application of the residue theorem results in
\begin{widetext}
\begin{align}
	I_{m\leq n,a}=&\,\dfrac{\E{-\frac{a^2}{2w_0^2}}}{\sqrt{2^{m+n}\,m!\,n!}}\!\!\!\!\!\!\!\!\!\!\underbrace{\left.\left(\dfrac{\partial}{\partial z}\right)^m\left[\E{-\frac{\sqrt{2}a}{w_0}z}\left(\dfrac{\sqrt{2}a}{w_0}+2z\right)^n\right]\right|_{z=0}}_{\qquad\qquad\qquad\qquad\qquad\qquad\qquad\qquad\qquad\qquad\!\displaystyle\mathclap{=\sum^m_{l=0}\binom{m}{l}\left(-\dfrac{\sqrt{2}a}{w_0}\right)^{m-l}\!\!\!\dfrac{n!}{(n-l)!}\,2^l\left(\dfrac{\sqrt{2}a}{w_0}\right)^{n-l}}}\nonumber\\
	=&\,\dfrac{\E{-\frac{a^2}{2w_0^2}}}{\sqrt{m!\,n!}}\left(\dfrac{a}{w_0}\right)^{m+n}\sum^m_{l=0}(-1)^{m-l}\left(\dfrac{w_0}{a}\right)^{2l}\binom{m}{l}\dfrac{n!}{(n-l)!}=\dfrac{\E{-\frac{a^2}{2w_0^2}}}{\sqrt{m!\,n!}}(-1)^m\left(\dfrac{a}{w_0}\right)^{n-m}\KUMMERU{-m}{1+m-n}{a^2/w_0^2}\,.
\end{align}
\end{widetext}
Here, $\KUMMERU{\cdot}{\cdot}{\cdot}$ is the Kummer confluent hypergeometric function. The second line follows from Leibniz's rule of differentiation and the identity
\begin{equation}
	\left(\dfrac{\D}{\D x}\right)^l\,x^n=\dfrac{n!}{(n-l)!}\,x^{n-l}
\end{equation}
for $l\leq n$. 

In order to complete the calculation, we need the expression for $I_{m>n,a}$. Without repeating the above procedures, we simply note that Eq.~\eqref{eq:I_m_n_a} has the equivalent form
\begin{align}
	I_{m\leq n,a}\equiv&\,\int\D x'\,u_{m}(x'-a)\,u_{n}(x')\nonumber\\
	=&\,\int\D x'\,u_{n}(x'+a)\,u_{m}(x')\equiv I_{m>n,a}
\end{align}
\emph{via} a trivial variable substitution and reordering of terms. This time, we may proceed to carry out the $z$ integral first, followed by the $z'$ one, where the displacement is now $-a$. This equivalence allows us to immediately state the answer
\begin{align}
	I_{m>n,a}=&\,\dfrac{\E{-\frac{a^2}{2w_0^2}}}{\sqrt{m!\,n!}}(-1)^m\left(\dfrac{a}{w_0}\right)^{m-n}\nonumber\\
	&\,\times\KUMMERU{-n}{1+n-m}{a^2/w_0^2}
\end{align}
by simply interchanging $m$ and $n$, and replacing $-a$ for $a$. Note that the phase factor retains the exponent $m$. Finally, introducing the shorthand notations $n_>=\max\{m,n\}$ and $n_<=\min\{m,n\}$ allows us to write down the compact form
\begin{align}
	I_{m,n,a}=&\,\dfrac{\E{-\frac{a^2}{2w_0^2}}}{\sqrt{m!\,n!}}(-1)^m\left(\dfrac{a}{w_0}\right)^{n_>-n_<}\nonumber\\
	&\,\times\KUMMERU{-n_<}{1+n_>-n_<}{a^2/w_0^2}\,.
\end{align}

\newpage
\onecolumngrid

\section*{Appendix tables and figures}

\begin{table*}[htp]
	\begin{tabular}{rcrrrr}
		$d$ & {\bf data type} & $N$ & $m$ & {\bf bs/loss (ICCNet)} & {\bf bs/loss (FidNet)} \\
		\cline{1-6}\\[-2ex]
		\cline{1-6}\\[-1ex]
		4 & random projectors& 500 & 10000 & 256/MAE & 1024/MSE \\
		4 & ACT bases & 5000 & 10000 & 256/MAE & 1024/MSE \\
		6 & ACT bases & 5000& 10000 & 256/MAE & 1024/MSE \\
		9 & ACT bases & 5000& 10000 & 256/MAE & 1024/MSE \\
		16 & Haar bases & 1000 & 5000 & 256/MAE & 1024/MSE \\
		16 & Haar bases & Inf & 5000 & 256/MAE & 1024/MSE \\
		16 & ACT bases & 1000 & 5000 & 256/MAE & 1024/MSE \\
		16 & ACT bases & Inf & 5000 & 256/MAE & 1024/MSE \\
		32 & Haar bases & 1000 & 1000 & 64/MAE & 512/MSE\\
		64 & Haar bases & 1000 & 1000 & 64/MAE & 512/MSE\\
		\cline{1-6}\\[-2ex]
		\cline{1-6}
	\end{tabular}
	\caption{\label{tab:data}A table of the simulated training data types, number of copies~$N$ per basis or projector, number of training datasets~$m$ for each value of $K$ or $L$, and the batch size~(bs) and loss function~(loss) used for training the respective Nets. For the first data type of $d=4$, the random projectors are chosen from the fixed set defined by \mbox{Eq.~(3.2)} and the 16 measurement angles listed in Fig.~11 in the main text.}
\end{table*}

\begin{table*}[htp]
	\begin{tabular}[t]{llllrccccccccc}
		& & & &  & \multicolumn{9}{c}{$K$}\\
		& & & & {$\bm{(d=16)}$} & {\bf2} & {\bf3} & {\bf4} & {\bf5} & {\bf6} & {\bf7} & {\bf8} & {\bf9} & {\bf10} \\
		\cline{2-14}\\[-2ex]
		\cline{2-14}\\[-1ex]
		& & & & {\bf act. avg.} & 0.20 & 0.80 & 4.26 & 6.42 & 6.93 & 7.41 & 7.53 & 7.83 & 7.84\\
		& & & & {\bf pred. avg.} & 0.17 & 0.65 & 3.98 & 5.64 & 6.66 & 7.00 & 8.09 & 8.44 & 8.54\\
		& & & \begin{rotate}{90}\,\,$r=1$\end{rotate} & {\bf avg. mae} & 0.05 & 0.23 & 1.35 & 0.89 & 0.64 & 0.61 & 0.62 & 0.68 & 0.76\\[1ex]
		& & & & {\bf act. avg.} & 0.10 & 0.25 & 0.46 & 0.84 & 2.2 & 5.18 & 6.43 & 6.95 & 7.33\\
		& & & & {\bf pred. avg.} & 0.10 & 0.25 & 0.40 & 1.24 & 2.82 & 4.5 & 5.37 & 7.07 & 7.57\\
		& & & \begin{rotate}{90}\,\,$r=2$\end{rotate} & {\bf avg. mae} & 0.02 & 0.05 & 0.10 & 0.64 & 1.42 & 1.23 & 1.22 & 0.69 & 0.66\\[1ex]
		& & & & {\bf act. avg.} & 0.08 & 0.18 & 0.31 & 0.47 & 0.72 & 1.12 & 2.67 & 5.31 & 6.34\\
		\begin{rotate}{90}Haar bases ($N=$Inf)\end{rotate} & & & & {\bf pred. avg.} & 0.08 & 0.18 & 0.29 & 0.51 & 0.86 & 2.36 & 3.97 & 5.04 & 5.81\\
		& & & \begin{rotate}{90}\,\,$r=3$\end{rotate} & {\bf avg. mae} & 0.02 & 0.02 & 0.06 & 0.15 & 0.26 & 1.35 & 1.57 & 0.95 & 1.11\\[1ex]
		& & & & {\bf act. avg.} & 0.23 & 1.76 & 4.14 & 4.65 & 5.44 & 6.12 & 7.50 & 8.10 & 8.39\\
		& & & & {\bf pred. avg.} & 0.17 & 0.64 & 3.98 & 5.58 & 6.72 & 7.02 & 8.09 & 8.47 & 8.56\\
		& & & \begin{rotate}{90}\,\,$r=1$\end{rotate} & {\bf avg. mae} & 0.07 & 1.17 & 0.46 & 1.07 & 1.48 & 1.42 & 0.76 & 0.59 & 0.48\\[1ex]
		& & & & {\bf act. avg.} & 0.11 & 0.28 & 0.72 & 3.32 & 4.35 & 4.53 & 4.84 & 6.13 & 7.12\\
		& & & & {\bf pred. avg.} & 0.10 & 0.27 & 0.68 & 1.58 & 3.28 & 4.86 & 5.73 & 7.14 & 7.63\\
		& & & \begin{rotate}{90}\,\,$r=2$\end{rotate} & {\bf avg. mae} & 0.02 & 0.07 & 0.48 & 1.86 & 1.35 & 0.62 & 1.01 & 1.29 & 1.17\\[1ex]
		& & & & {\bf act. avg.} & 0.09 & 0.20 & 0.35 & 0.65 & 2.51 & 4.33 & 4.49 & 4.71 & 5.08\\
		\begin{rotate}{90}Haar bases ($N=5000$)\end{rotate} & & & & {\bf pred. avg.} & 0.08 &  0.19 & 0.29 & 0.55 & 1.09 & 2.86 & 4.16 & 5.30 & 5.82\\
		& & & \begin{rotate}{90}\,\,$r=3$\end{rotate} & {\bf avg. mae} & 0.02 & 0.03 & 0.07 & 0.26 & 1.56 & 1.48 & 0.50 & 0.78 & 0.99\\[1ex]
		& & & & {\bf act. avg.} & 0.19 & 0.54 & 6.74 & 8.25 & 8.1 & 8.14 & 8.12 & 8.26 & 8.30\\
		& & & & {\bf pred. avg.} & 0.17 & 0.69 & 6.64 & 6.56 & 7.29 & 7.87 & 8.18 & 8.19 & 8.32\\
		& & & \begin{rotate}{90}\,\,$r=1$\end{rotate} & {\bf avg. mae} & 0.05 & 0.17 & 1.12 & 1.72 & 0.86 & 0.69 & 0.66 & 0.61 & 0.73\\[1ex]
		& & & & {\bf act. avg.} & 0.10 & 0.24 & 0.46 & 0.95 & 6.12 & 7.04 & 7.39 & 7.62 & 7.73\\
		& & & & {\bf pred. avg.} & 0.10 & 0.24 & 0.42 & 1.13 & 5.78 & 6.55 & 6.91 & 7.53 & 7.64\\
		& & & \begin{rotate}{90}\,\,$r=2$\end{rotate} & {\bf avg. mae} & 0.02 & 0.05 & 0.16 & 0.49 & 0.99 & 0.81 & 0.75 & 0.37 & 0.35\\[1ex]
		& & & & {\bf act. avg.} & 0.08 & 0.17 & 0.30 & 0.47 & 0.69 & 1.96 & 5.62 & 6.27 & 6.76\\
		\begin{rotate}{90}ACT bases ($N=$Inf)\end{rotate} & & & & {\bf pred. avg.} & 0.08 & 0.17 & 0.25 & 0.45 & 1.19 & 3.38 & 5.19 & 6.60 & 6.70\\
		& & & \begin{rotate}{90}\,\,$r=3$\end{rotate} & {\bf avg. mae} & 0.01 & 0.02 & 0.06 & 0.11 & 0.59 & 1.62 & 0.99 & 0.87 & 0.60\\[1ex]
		& & & & {\bf act. avg.} & 0.25 & 1.36 & 4.50 & 4.98 & 6.34 & 6.73 & 7.65 & 7.8 & 8.23\\
		& & & & {\bf pred. avg.} & 0.18 & 0.68 & 5.02 & 6.31 & 7.19 & 7.75 & 8.14 & 8.15 & 8.30\\
		& & & \begin{rotate}{90}\,\,$r=1$\end{rotate} & {\bf avg. mae} & 0.10 & 0.78 & 0.53 & 1.38 & 1.05 & 1.23 & 0.62 & 0.53 & 0.49\\[1ex]
		& & & & {\bf act. avg.} & 0.10 & 0.26 & 0.68 & 2.65 & 4.32 & 4.52 & 4.69 & 5.40 & 5.88\\
		& & & & {\bf pred. avg.} & 0.10 & 0.26 & 0.67 & 1.48 & 4.72 & 4.94 & 4.77 & 5.64 & 5.62\\
		& & & \begin{rotate}{90}\,\,$r=2$\end{rotate} & {\bf avg. mae} & 0.02 & 0.07 & 0.39 & 1.46 & 0.77 & 0.49 & 0.24 & 0.74 & 0.90\\[1ex]
		& & & & {\bf act. avg.} & 0.08 & 0.19 & 0.33 & 0.68 & 2.43 & 4.18 & 4.33 & 4.47 & 4.66\\
		\begin{rotate}{90}ACT bases ($N=5000$)\end{rotate} & & & & {\bf pred. avg.} & 0.08 &  0.18 & 0.26 & 0.43 & 1.04 & 3.13 & 4.52 & 4.44 & 4.55\\
		& & & \begin{rotate}{90}\,\,$r=3$\end{rotate} & {\bf avg. mae} & 0.02 & 0.03 & 0.08 & 0.30 & 1.62 & 1.15 & 0.28 & 0.13 & 0.18\\[1ex]
		\cline{2-14}\\[-2ex]
		\cline{2-14}
	\end{tabular}~~\begin{tabular}[t]{ccccccccclrrrr}
		\multicolumn{9}{c}{$K$} & & &  &&\\
		{\bf2} & {\bf3} & {\bf4} & {\bf5} & {\bf6} & {\bf7} & {\bf8} & {\bf9} & {\bf10} & & & & &\\
		\cline{1-12}\\[-2ex]
		\cline{1-12}\\[-1ex]
		0.32 & 1.55 & 4.57 & 6.10 & 5.87 & 5.83 & 5.85 & 5.87 & 5.82 & \begin{rotate}{270}\!\!$d=4$\end{rotate} & & & \begin{rotate}{270}Hermite-Gaussian modes\end{rotate}&\\
		0.22 & 1.75 & 5.21 & 6.18 & 6.63 & 6.65 & 6.70 & 6.61 & 6.65 & & & &&\\
		0.15 & 1.00 & 1.49 & 0.54 & 1.03 & 0.86 & 0.88 & 0.81 & 0.88 & & & & &\\[1ex]
		0.24 & 1.09 & 3.51 & 6.63 & 6.75 & 6.95 & 6.79 & 6.72 & 6.86 & \begin{rotate}{270}\!\!$d=6$\end{rotate} & & &&\\
		0.21 & 1.27 & 4.51 & 5.30 & 5.56 & 6.34 & 6.33 & 6.50 & 6.50 & & & &&\\
		0.08 & 0.46 & 1.79 & 2.42 & 1.39 & 1.02 & 0.93 & 0.87 & 0.86 & & & &&\\[1ex]
		0.22 & 0.87 & 2.75 & 5.95 & 6.63 & 6.09 & 6.38 & 6.20 & 6.28 & \begin{rotate}{270}\!\!$d=9$\end{rotate} & & & &\\
		0.14 & 1.81 & 5.23 & 5.65 & 6.27 & 6.87 & 7.68 & 5.69 & 6.14 & & & & &\\
		0.10 & 1.17 & 3.03 & 2.24 & 1.04 & 1.35 & 1.48 & 1.02 & 1.13 & & & &&\\[1ex]
		
		0.32 & 1.55 & 4.57 & 6.10 & 5.87 & 5.83 & 5.85 & 5.87 & 5.82 & \begin{rotate}{270}\!\!$d=4$\end{rotate} & & & \begin{rotate}{270}Hermite-Gaussian modes\end{rotate}&\\
		0.31 & 1.39 & 5.45 & 6.44 & 6.78 & 6.76 & 6.74 & 6.67 & 6.70 & & & &&\,\,\,\,\begin{rotate}{270}\,\,\,\,\,(noise-trained)\end{rotate}\\
		0.15& 1.02& 1.61& 0.53& 0.95& 0.96& 0.92& 0.87& 0.94 & & & & &\\[1ex]
		0.24 & 1.09 & 3.51 & 6.63 & 6.75 & 6.95 & 6.79 & 6.72 & 6.86 & \begin{rotate}{270}\!\!$d=6$\end{rotate} & & &&\\
		0.22 & 1.61 & 5.27 & 6.40 & 5.75 & 6.37 & 6.48 & 6.53 & 6.47 & & & &&\\
		0.10 & 0.83 &2.47& 2.16& 1.41& 1.16& 0.95& 0.88& 0.92 & & & &&\\[1ex]
		0.22 & 0.87 & 2.75 & 5.95 & 6.63 & 6.09 & 6.38 & 6.20 & 6.28 & \begin{rotate}{270}\!\!$d=9$\end{rotate} & & & &\\
		0.32 & 1.30 & 5.41 & 5.46 & 6.42 & 6.70 & 7.42 & 6.27 & 6.23 & & & & &\\
		0.14 &0.91 &2.68& 2.19 &1.27& 1.16& 1.52& 1.07& 1.05 & & & &&\\[1ex]
		\cline{1-12}\\[-2ex]
		\cline{1-12}
	\end{tabular}\\	
	\begin{tabular}[t]{lllllrcccccccccccccc}
		&& & & & &\multicolumn{14}{c}{$L$}\\
		&& & & & {$\bm{(d=4)}$} &{\bf2} & {\bf3} & {\bf4} & {\bf5} & {\bf6} & {\bf7} & {\bf8} & {\bf9} & {\bf10} & {\bf11} & {\bf12} & {\bf13} & {\bf14} & {\bf15}\\
		\cline{3-20}\\[-2ex]
		\cline{3-20}\\[-1ex]			
		&& & & & {\bf act. avg.} & 0.03 & 0.10 & 0.16 & 0.21 & 0.35 & 0.80 & 1.93 & 3.09 & 4.27 & 4.37 & 7.05 & 8.28 & 8.71 & 8.51 \\
		&& & & & {\bf pred. avg.} & 0.04 & 0.07 & 0.12&  0.20& 0.29& 0.36 & 0.47 & 0.78 & 0.72& 1.28& 3.20& 6.70& 7.25& 5.02\\
		&& & & \begin{rotate}{90}state 1\end{rotate}& {\bf avg. mae} &0.03& 0.07& 0.08& 0.07& 0.12& 0.50 & 1.53& 2.44& 3.56& 3.17& 4.08& 1.93& 1.52& 3.49\\[1ex]
		&&  & & & {\bf act. avg.} & 0.04 & 0.09 & 0.15 & 0.23 & 0.31 & 0.41 & 0.73 & 1.74 & 2.72 & 3.84 & 7.05 & 7.28 & 7.88 & 8.53 \\
		&& & & & {\bf pred. avg.} & 0.03 & 0.07 & 0.12 & 0.19 & 0.27 & 0.36 & 0.45 & 0.71 & 0.88 & 1.51 & 4.70& 6.30& 7.31& 6.55\\
		&& & & \begin{rotate}{90}state 2\end{rotate}& {\bf avg. mae} & 0.02& 0.05& 0.06& 0.07& 0.10 & 0.11& 0.34 &1.09& 1.84& 2.34& 2.81& 2.24& 0.61& 1.98\\[1ex]
		&\begin{rotate}{90}\!\!\!\!\!Three-photon states\end{rotate}& & & & {\bf act. avg.} & 0.05 & 0.06 & 0.10 & 0.15 & 0.21 & 0.26 & 0.34 & 0.47 & 0.87 & 1.50 & 4.41 & 5.56 & 5.53 & 5.42\\
		&& & &  & {\bf pred. avg.} & 0.04 &0.07& 0.10& 0.16& 0.24& 0.29& 0.39& 0.46& 0.65 & 0.90 & 1.17& 1.48& 2.59 & 5.01 \\
		&& & & \begin{rotate}{90}state 3\end{rotate}& {\bf avg. mae} & 0.02& 0.05& 0.04& 0.06& 0.08& 0.09& 0.11& 0.11& 0.35& 0.85& 3.41& 4.27& 2.95& 0.42\\[1ex]
		
		&& & & & {\bf act. avg.} & 0.03 & 0.10 & 0.16 & 0.21 & 0.35 & 0.80 & 1.93 & 3.09 & 4.27 & 4.37 & 7.05 & 8.28 & 8.71 & 8.51 \\
		&& & & & {\bf pred. avg.} & 0.04 & 0.09 & 0.13 & 0.22 & 0.35 & 0.46 & 0.65 & 2.30 & 3.64 & 4.53 & 7.91 & 8.23 & 8.66 & 8.51\\
		&& & & \begin{rotate}{90}state 1\end{rotate}& {\bf avg. mae} &0.03 & 0.05 & 0.08 & 0.08 & 0.13 & 0.48 & 1.45 & 1.78 & 1.58 & 0.97 & 1.36 & 0.61 & 0.34 & 0.28\\[1ex]
		&&  & & & {\bf act. avg.} & 0.04 & 0.09 & 0.15 & 0.23 & 0.31 & 0.41 & 0.73 & 1.74 & 2.72 & 3.84 & 7.05 & 7.28 & 7.88 & 8.53 \\
		&& & & & {\bf pred. avg.} & 0.04 & 0.09 & 0.13 & 0.20 & 0.26 & 0.37 & 0.52 & 0.61 & 1.13 & 2.59 & 6.61 & 6.92 & 8.15 & 8.50\\
		&& & & \begin{rotate}{90}state 2\end{rotate}& {\bf avg. mae} & 0.01 & 0.03 & 0.05 & 0.06 & 0.11 & 0.13 & 0.28 & 1.15 & 1.75 & 1.69 & 1.92 & 2.23 & 1.13 & 0.11\\[1ex]
		\begin{rotate}{90}\!(noise-trained)\end{rotate}\,\,\,\,\,&\begin{rotate}{90}\!\!\!\!\!Three-photon states\end{rotate}& & & & {\bf act. avg.} & 0.05 & 0.06 & 0.10 & 0.15 & 0.21 & 0.26 & 0.34 & 0.47 & 0.87 & 1.50 & 4.41 & 5.56 & 5.53 & 5.42\\
		&& & &  & {\bf pred. avg.} & 0.05 & 0.07 & 0.10 & 0.17 & 0.26 & 0.32 & 0.40 & 0.49 & 0.63 & 0.84 & 1.79 & 4.38 & 5.08 & 5.39 \\
		&& & & \begin{rotate}{90}state 3\end{rotate}& {\bf avg. mae} & 0.01 & 0.02 & 0.04 & 0.07 & 0.07 & 0.09 & 0.11 & 0.12 & 0.39 & 0.82 & 3.79 & 3.11 & 0.46 & 0.07\\[1ex]		
		\cline{3-20}\\[-2ex]
		\cline{3-20}
	\end{tabular}
	\caption{\label{tab:iccnet}Table of the actual averages, predicted ones and average MAE of $-\log_{10} s_{\textsc{cvx}}$ that are obtained from all trained ICCNet models.}
\end{table*}

\begin{table*}[htp]
	\begin{tabular}[t]{llllrcccccccccc}
		& & & &  & \multicolumn{10}{c}{$K$}\\
		& & & & {$\bm{(d=16)}$} & {\bf1} & {\bf2} & {\bf3} & {\bf4} & {\bf5} & {\bf6} & {\bf7} & {\bf8} & {\bf9} & {\bf10} \\
		\cline{2-15}\\[-2ex]
		\cline{2-15}\\[-1ex]
		& & & & {\bf act. avg.} & 0.11 & 0.21 & 0.52 & 0.98 & 1.00 & 1.00 &  1.00 &  1.00 &  1.00 &  1.00\\
		& & & & {\bf pred. avg.} & 0.11 & 0.21& 0.52& 0.91& 0.98& 0.97& 0.99& 0.99& 0.99& 0.99\\
		& & & \begin{rotate}{90}\,\,$r=1$\end{rotate} & {\bf avg. mae} & 0.01& 0.03 &0.10&  0.08& 0.02& 0.03& 0.01& 0.01& 0.01& 0.01\\[1ex]
		& & & & {\bf act. avg.} & 0.17& 0.24 &0.31 &0.43 &0.64 &0.89 &1.00&   1.00&   1.00&   1.00\\
		& & & & {\bf pred. avg.} & 0.18& 0.25& 0.33& 0.46& 0.67& 0.85 &0.92& 0.94& 0.95& 0.95\\
		& & & \begin{rotate}{90}\,\,$r=2$\end{rotate} & {\bf avg. mae} & 0.01& 0.02& 0.04& 0.05& 0.08& 0.08& 0.08 &0.06 &0.05& 0.05\\[1ex]
		& & & & {\bf act. avg.} & 0.22& 0.27 &0.33 &0.41& 0.51& 0.64& 0.81& 0.95& 1.00&   1.00\\
		\begin{rotate}{90}Haar bases ($N=$Inf)\end{rotate} & & & & {\bf pred. avg.} & 0.22 &0.28& 0.33& 0.42& 0.52& 0.67& 0.79& 0.88& 0.92& 0.93\\
		& & & \begin{rotate}{90}\,\,$r=3$\end{rotate} & {\bf avg. mae} & 0.01 &0.02 &0.02& 0.03& 0.04& 0.06& 0.06& 0.08& 0.08 &0.07\\[1ex]
		& & & & {\bf act. avg.} & 0.12& 0.23& 0.58& 0.84& 0.91& 0.93& 0.95& 0.95& 0.96& 0.96\\
		& & & & {\bf pred. avg.} & 0.11& 0.21& 0.52& 0.91& 0.98& 0.97& 0.99 &0.99& 0.99& 0.99\\
		& & & \begin{rotate}{90}\,\,$r=1$\end{rotate} & {\bf avg. mae} & 0.01& 0.04& 0.15& 0.08& 0.07& 0.04& 0.04& 0.04& 0.03& 0.03\\[1ex]
		& & & & {\bf act. avg.} & 0.18 &0.24 &0.33 &0.47 &0.69 &0.81 &0.85 &0.88 &0.90 & 0.92\\
		& & & & {\bf pred. avg.} & 0.18 &0.25 &0.34 &0.47 &0.68 &0.84 &0.91 &0.95 &0.95 &0.95\\
		& & & \begin{rotate}{90}\,\,$r=2$\end{rotate} & {\bf avg. mae} & 0.01& 0.03& 0.03& 0.06& 0.08& 0.05& 0.06& 0.06& 0.04& 0.03\\[1ex]
		& & & & {\bf act. avg.} & 0.23& 0.28 &0.34 &0.42& 0.54 &0.69& 0.78& 0.82& 0.85& 0.87\\
		\begin{rotate}{90}Haar bases ($N=5000$)\end{rotate} & & & & {\bf pred. avg.} & 0.22& 0.29 &0.33& 0.42& 0.53& 0.69& 0.80 & 0.89 &0.92& 0.93\\
		& & & \begin{rotate}{90}\,\,$r=3$\end{rotate} & {\bf avg. mae} & 0.01& 0.02& 0.02& 0.03& 0.05& 0.08& 0.05& 0.07& 0.07& 0.06\\[1ex]
		& & & & {\bf act. avg.} & 0.12& 0.22& 0.51& 0.95& 1.00&   1.00&   1.00&   1.00&   1.00&   1.00\\
		& & & & {\bf pred. avg.} & 0.11 & 0.21& 0.52& 0.96& 0.99& 0.99& 1.00&   1.00&   1.00&   1.00 \\
		& & & \begin{rotate}{90}\,\,$r=1$\end{rotate} & {\bf avg. mae} & 0.01 &0.04& 0.12& 0.05& 0.01 &0.01 &0.00&   0.00&   0.00&   0.00\\[1ex]
		& & & & {\bf act. avg.} & 0.17& 0.24& 0.33& 0.47& 0.75& 0.99& 1.00&   1.00&   1.00&   1.00\\
		& & & & {\bf pred. avg.} & 0.18& 0.25& 0.34& 0.47 &0.77& 0.98& 0.99& 1.00&   1.00&   1.00\\
		& & & \begin{rotate}{90}\,\,$r=2$\end{rotate} & {\bf avg. mae} & 0.01 &0.02 &0.04 &0.06& 0.06& 0.02& 0.01& 0.00&   0.00&   0.00\\[1ex]
		& & & & {\bf act. avg.} & 0.23& 0.28 &0.34 &0.43& 0.53& 0.66& 0.93& 1.00&   1.00&   1.00\\
		\begin{rotate}{90}ACT bases ($N=$Inf)\end{rotate} & & & & {\bf pred. avg.} & 0.22 &0.29& 0.34& 0.42& 0.53 &0.66 &0.92& 0.99& 1.00&   1.00\\
		& & & \begin{rotate}{90}\,\,$r=3$\end{rotate} & {\bf avg. mae} &0.01& 0.02& 0.02& 0.03& 0.04& 0.05& 0.02& 0.01& 0.00&   0.00\\[1ex]
		& & & & {\bf act. avg.} & 0.12 &0.21& 0.51& 0.87& 0.96& 0.98& 0.98& 0.98& 0.99& 0.99\\
		& & & & {\bf pred. avg.} & 0.11& 0.21& 0.53 &0.96 &0.99 &0.99 &0.99 &1.00&   1.00&   1.00 \\
		& & & \begin{rotate}{90}\,\,$r=1$\end{rotate} & {\bf avg. mae} & 0.01 &0.03 &0.12 &0.09& 0.03& 0.01& 0.02& 0.01 &0.01 &0.01\\[1ex]
		& & & & {\bf act. avg.} & 0.18& 0.24& 0.34& 0.52 &0.72& 0.84 &0.89 &0.92 &0.94& 0.95\\
		& & & & {\bf pred. avg.} & 0.18& 0.25& 0.35 &0.48 &0.73& 0.87 &0.94 &0.95 &0.95& 0.97\\
		& & & \begin{rotate}{90}\,\,$r=2$\end{rotate} & {\bf avg. mae} & 0.01 &0.03 &0.05 &0.09& 0.06& 0.05& 0.05& 0.03& 0.02& 0.03\\[1ex]
		& & & & {\bf act. avg.} & 0.23& 0.28& 0.34& 0.42& 0.54 &0.67 &0.76 &0.81& 0.85& 0.88\\
		\begin{rotate}{90}ACT bases ($N=5000$)\end{rotate} & & & & {\bf pred. avg.} & 0.22& 0.29& 0.34 &0.42& 0.53& 0.67& 0.76 &0.82& 0.85& 0.88\\
		& & & \begin{rotate}{90}\,\,$r=3$\end{rotate} & {\bf avg. mae} & 0.01 &0.02& 0.03 &0.04& 0.07 &0.05 &0.03& 0.04& 0.04 &0.03\\[1ex]
		\cline{2-15}\\[-2ex]
		\cline{2-15}
	\end{tabular}~~\begin{tabular}[t]{cccccccccclrrrrrr}
		\multicolumn{10}{c}{$K$} & & &  & &&&\\
		{\bf1} & {\bf2} & {\bf3} & {\bf4} & {\bf5} & {\bf6} & {\bf7} & {\bf8} & {\bf9} & {\bf10} & & & & &&&\\
		\cline{1-13}\\[-2ex]
		\cline{1-13}\\[-1ex]
		0.48 & 0.65 & 0.83 & 0.86 & 0.92 & 0.95 & 0.96 & 0.97 & 0.98 & 0.98 & \begin{rotate}{270}\!\!$d=4$\end{rotate} & & & \begin{rotate}{270}Hermite-Gaussian modes\end{rotate}&&&\\
		0.50 & 0.72 & 0.90 & 0.95 & 0.98 & 0.98 & 0.97 & 0.98 & 0.97 & 0.97 & & & &&&&\\
		0.07 & 0.16 & 0.11 & 0.11 & 0.05 & 0.03 & 0.03 & 0.03 & 0.04 & 0.05 & & & &&&& \\[1ex]
		0.29 & 0.49 & 0.75 & 0.81 & 0.90 & 0.94 & 0.96 & 0.97 & 0.97 & 0.98 & \begin{rotate}{270}\!\!$d=6$\end{rotate} & & &&&&\\
		0.33 & 0.49 & 0.83 & 0.92 & 0.97 & 0.92 & 0.98 & 0.95 & 0.95 & 0.98 & & & &&&&\\
		0.08 & 0.17 & 0.14 & 0.15 & 0.07 & 0.06 & 0.03 & 0.05 & 0.05 & 0.02 & & & &&&&\\[1ex]
		0.32 & 0.53 & 0.68 & 0.79 & 0.89 & 0.93 & 0.95 & 0.96 & 0.97 & 0.97 & \begin{rotate}{270}\!\!$d=9$\end{rotate} & & & &&&\\
		0.20 & 0.50 & 0.75 & 0.86 & 0.90 & 0.97 & 0.90 & 0.87 & 0.90 & 0.97 & & & & &&&\\
		0.14 & 0.14 & 0.16 & 0.14 & 0.08 & 0.05 & 0.07 & 0.11 & 0.07 & 0.02 & & & &&&&\\[1ex]
		
		0.48 & 0.65 & 0.83 & 0.86 & 0.92 & 0.95 & 0.96 & 0.97 & 0.98 & 0.98 & \begin{rotate}{270}\!\!$d=4$\end{rotate} & & & \begin{rotate}{270}Hermite-Gaussian modes\end{rotate}&&&\\
		0.49 & 0.64 & 0.83 & 0.94 & 0.97 & 0.97 & 0.98 & 0.98 & 0.97 & 0.97 & & & &&&&\begin{rotate}{270}\,\,\,\,\,(noise-trained)\end{rotate}\\
		0.05 & 0.17 & 0.14 & 0.10 & 0.05 & 0.05 & 0.03 & 0.03 & 0.04 & 0.04 & & & &&&& \\[1ex]
		0.29 & 0.49 & 0.75 & 0.81 & 0.90 & 0.94 & 0.96 & 0.97 & 0.97 & 0.98 & \begin{rotate}{270}\!\!$d=6$\end{rotate} & & &&&&\\
		0.27 & 0.45 & 0.75 & 0.90 & 0.94 & 0.94 & 0.97 & 0.96 & 0.96 & 0.96 & & & &&&&\\
		0.05 & 0.19 & 0.13 & 0.11 & 0.05 & 0.04 & 0.02 & 0.03 & 0.04 & 0.03 & & & &&&&\\[1ex]
		0.32 & 0.53 & 0.68 & 0.79 & 0.89 & 0.93 & 0.95 & 0.96 & 0.97 & 0.97 & \begin{rotate}{270}\!\!$d=9$\end{rotate} & & & &&&\\
		0.33 & 0.38 & 0.65 & 0.90 & 0.94 & 0.93 & 0.93 & 0.91 & 0.88 & 0.93 & & & & &&&\\
		0.04 & 0.20 & 0.14 & 0.12 & 0.05 & 0.03 & 0.04 & 0.07 & 0.09 & 0.05 & & & &&&&\\[1ex]		
		\cline{1-13}\\[-2ex]
		\cline{1-13}
	\end{tabular}\\	
	\begin{tabular}[t]{lllllrccccccccccccccc}
		&& & & & &\multicolumn{15}{c}{$L$}\\
		&& & & & {$\bm{(d=4)}$} &{\bf1} &{\bf2} & {\bf3} & {\bf4} & {\bf5} & {\bf6} & {\bf7} & {\bf8} & {\bf9} & {\bf10} & {\bf11} & {\bf12} & {\bf13} & {\bf14} & {\bf15}\\
		\cline{3-21}\\[-2ex]
		\cline{3-21}\\[-1ex]
		&& & & & {\bf act. avg.} & 0.38 & 0.44 & 0.49 & 0.47 & 0.47 & 0.46 & 0.52 & 0.70 & 0.78 & 0.86 & 0.89 & 0.89 & 0.91 & 0.92 & 0.93 \\
		&& & & & {\bf pred. avg.} & 0.34 & 0.40 & 0.37 & 0.46 & 0.48 & 0.49 & 0.58 & 0.64 & 0.77 & 0.83 & 0.89 & 0.86 & 0.88 & 0.91 & 0.92\\
		&& & & \begin{rotate}{90}state 1\end{rotate}& {\bf avg. mae} & 0.07 & 0.09 & 0.15 & 0.13 & 0.18 & 0.17 & 0.18 & 0.18 & 0.12 & 0.08 & 0.05 & 0.05 & 0.04 & 0.02 & 0.01\\[1ex]
		&&  & & & {\bf act. avg.} & 0.12 & 0.21 & 0.30 & 0.30 & 0.31 & 0.32 & 0.31 & 0.43 & 0.65 & 0.78 & 0.81 & 0.85 & 0.87 & 0.87 & 0.88 \\
		&& & & & {\bf pred. avg.} & 0.13 & 0.18 & 0.21 & 0.34 & 0.35 & 0.33 & 0.56 & 0.61 & 0.74 & 0.81 & 0.89 & 0.87 & 0.87 & 0.90 & 0.92\\
		&& & & \begin{rotate}{90}state 2\end{rotate}& {\bf avg. mae} & 0.05 & 0.11 & 0.13 & 0.12 & 0.16 & 0.15 & 0.27 & 0.26 & 0.17 & 0.10 & 0.09 & 0.04 & 0.03 & 0.03 & 0.04\\[1ex]
		&\begin{rotate}{90}\!\!\!\!\!Three-photon states\end{rotate}& & & & {\bf act. avg.} & 0.26 & 0.29 & 0.33 & 0.34 & 0.41 & 0.47 & 0.47 & 0.57 & 0.63 & 0.74 & 0.80 & 0.89 & 0.91 & 0.93 & 0.94\\
		&& & &  & {\bf pred. avg.} & 0.29 & 0.29 & 0.31 & 0.38 & 0.42 & 0.47 & 0.54 & 0.53 & 0.61 & 0.70 & 0.81 & 0.87 & 0.85 & 0.89 & 0.92 \\
		&& & & \begin{rotate}{90}state 3\end{rotate}& {\bf avg. mae} & 0.05 & 0.08 & 0.10 & 0.08 & 0.06 & 0.08 & 0.14 & 0.13 & 0.12 & 0.11 & 0.12 & 0.06 & 0.07 & 0.03 & 0.02\\[1ex]
		
		&& & & & {\bf act. avg.} & 0.38 & 0.44 & 0.49 & 0.47 & 0.47 & 0.46 & 0.52 & 0.70 & 0.78 & 0.86 & 0.89 & 0.89 & 0.91 & 0.92 & 0.93 \\
		&& & & & {\bf pred. avg.} & 0.37 & 0.41 & 0.43 & 0.42 & 0.43 & 0.43 & 0.53 & 0.61 & 0.74 & 0.81 & 0.85 & 0.88 & 0.92 & 0.94 & 0.94\\
		&& & & \begin{rotate}{90}state 1\end{rotate}& {\bf avg. mae} & 0.06 & 0.08 & 0.12 & 0.14  & 0.17 & 0.16 & 0.15 & 0.19 & 0.13 & 0.09 & 0.06 & 0.05 & 0.03 & 0.02 & 0.01\\[1ex]
		&&  & & & {\bf act. avg.} & 0.12 & 0.21 & 0.30 & 0.30 & 0.31 & 0.32 & 0.31 & 0.43 & 0.65 & 0.78 & 0.81 & 0.85 & 0.87 & 0.87 & 0.88 \\
		&& & & & {\bf pred. avg.} & 0.14 & 0.22 & 0.26 & 0.28 & 0.29 & 0.32 & 0.42 & 0.51 & 0.64 & 0.78 & 0.80 & 0.84 & 0.90 & 0.90 & 0.91\\
		&& & & \begin{rotate}{90}state 2\end{rotate}& {\bf avg. mae} & 0.06 & 0.08 & 0.11 & 0.12 & 0.14 & 0.14 & 0.19 & 0.20 & 0.16 & 0.09 & 0.08 & 0.05 & 0.03 & 0.03 & 0.03\\[1ex]
		\begin{rotate}{90}\!(noise-trained)\end{rotate}\,\,\,\,\,&\begin{rotate}{90}\!\!\!\!\!Three-photon states\end{rotate}& & & & {\bf act. avg.} & 0.26 & 0.29 & 0.33 & 0.34 & 0.41 & 0.47 & 0.47 & 0.57 & 0.63 & 0.74 & 0.80 & 0.89 & 0.91 & 0.93 & 0.94\\
		&& & &  & {\bf pred. avg.} & 0.26 & 0.28 & 0.32 & 0.36 & 0.40 & 0.46 & 0.49 & 0.56 & 0.61  &0.71 & 0.79 & 0.83  &0.89 & 0.92&0.93 \\
		&& & & \begin{rotate}{90}state 3\end{rotate}& {\bf avg. mae} & 0.04 &0.07& 0.08& 0.08 &0.08& 0.06& 0.10 & 0.13& 0.14& 0.11& 0.10& 0.07& 0.04& 0.01& 0.01\\[1ex]
		\cline{3-21}\\[-2ex]
		\cline{3-21}
	\end{tabular}
	\caption{\label{tab:fidnet}Table of the actual averages, predicted ones and average MAE of $\mathcal{F}$ that are obtained from all trained FidNet models.}
\end{table*}

\begin{figure*}[htp]
	\centering\includegraphics[width=1\columnwidth]{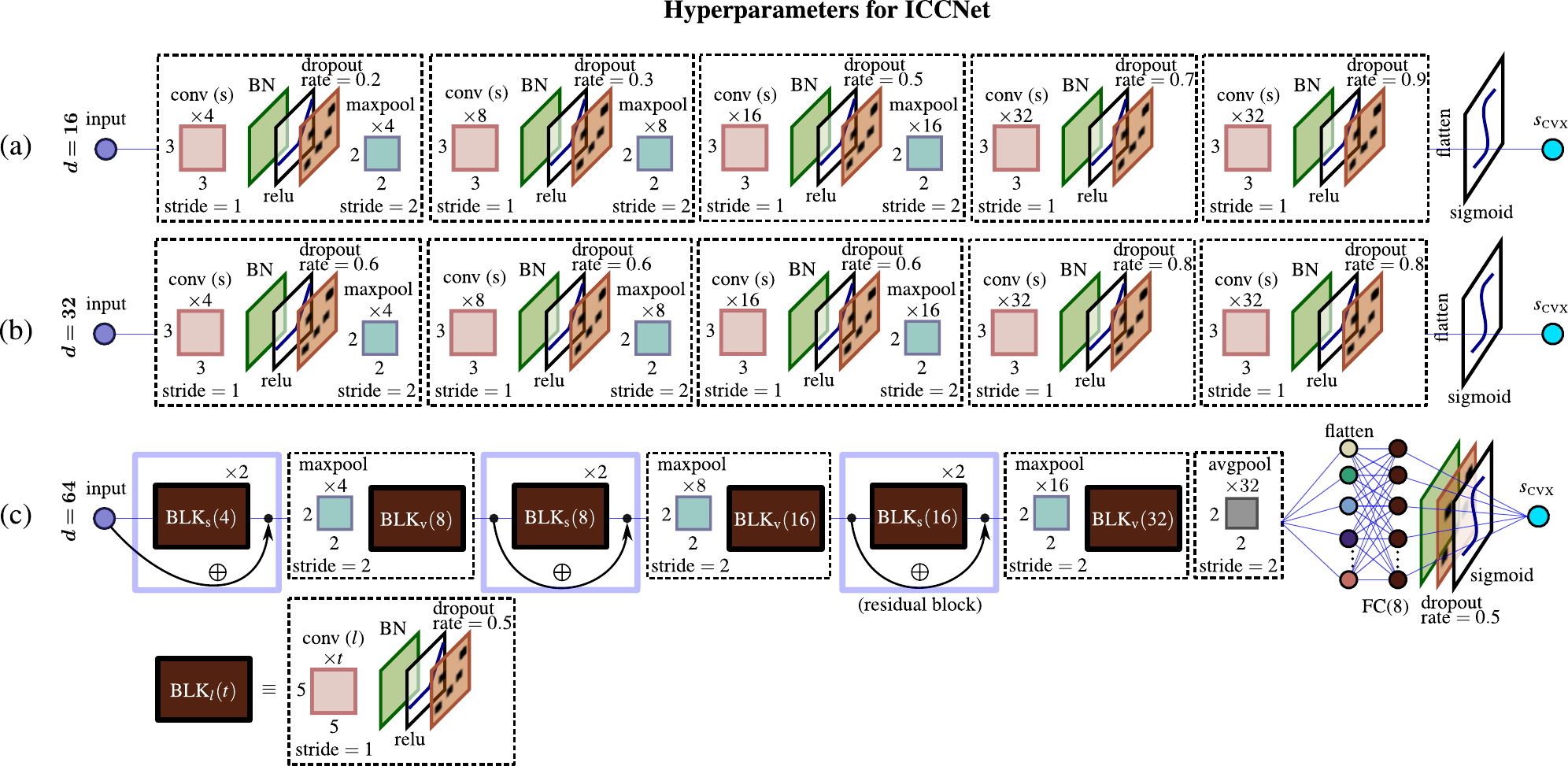}
	\caption{\label{fig:hyperparam_ICCNet} A list of hyperparameters for ICCNet (filter number, kernel or filter dimensions, stride length and dropout rate) used during training for various Hilbert-space dimensions. The notations conv~(s) and conv~(v) signifies whether zeros are padded to the output~(same) in order to maintain the same array dimensions after going through the convolution layer or not~(valid). For $d=4$, 6 and 9, network~(a) is adopted with the first two dropout rates set to zero. Additionally, for $d=4$, we change the max-pooling stride length to 1. Network~(c) consists of a deeper CNN layer with additional elements like an average-pooling layer and a fully-connected layer with 8 artificial neurons.}
\end{figure*}

\begin{figure*}[htp]
	\centering\includegraphics[width=1\columnwidth]{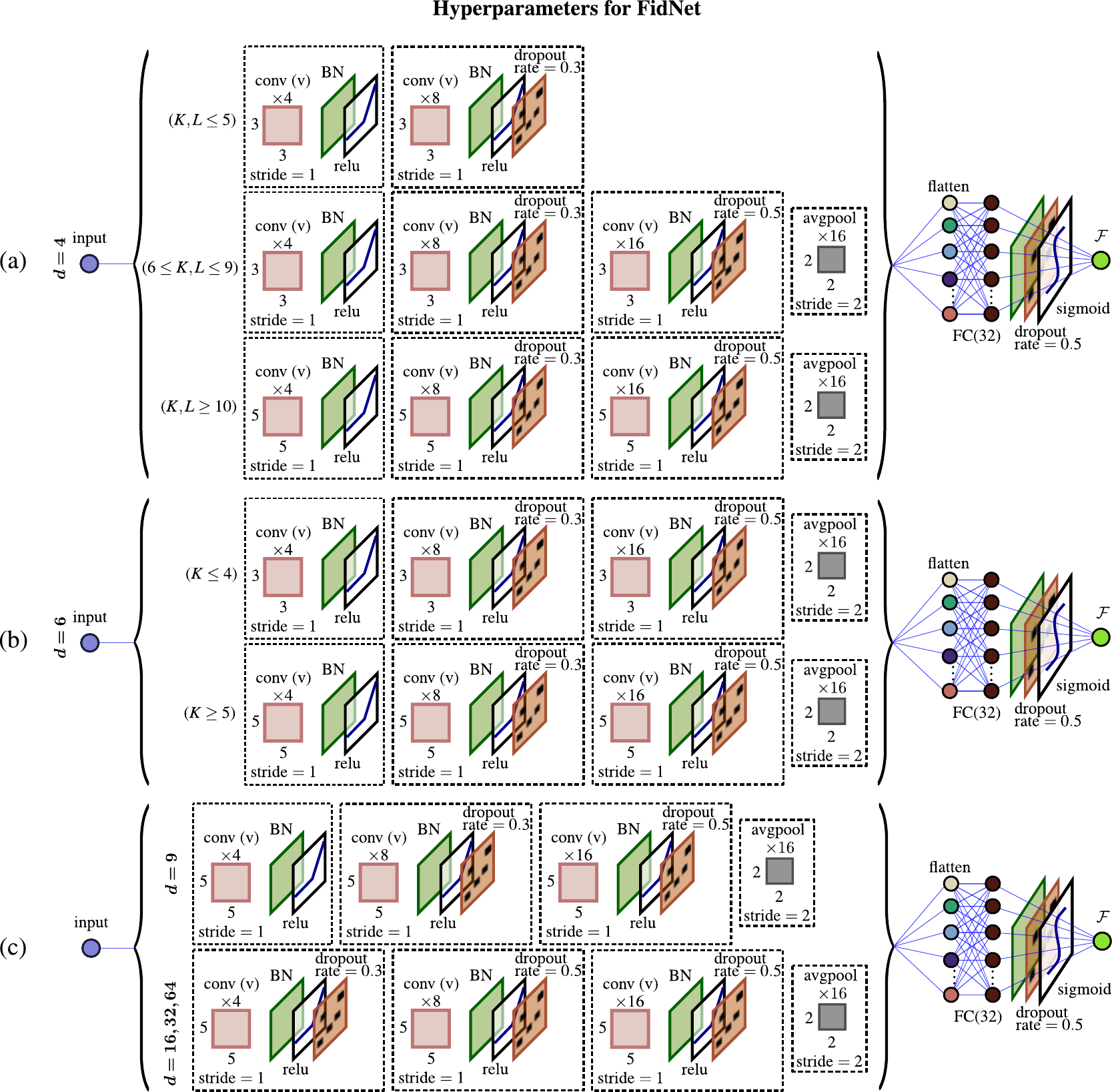}
	\caption{\label{fig:hyperparam_FidNet}The hyperparameter settings used to train FidNet and plot the figures in the main article. A single average-pooling layer is sufficient in all the FidNet architectures.}
\end{figure*}

\begin{figure*}[htp]
	\centering\includegraphics[width=0.8\columnwidth]{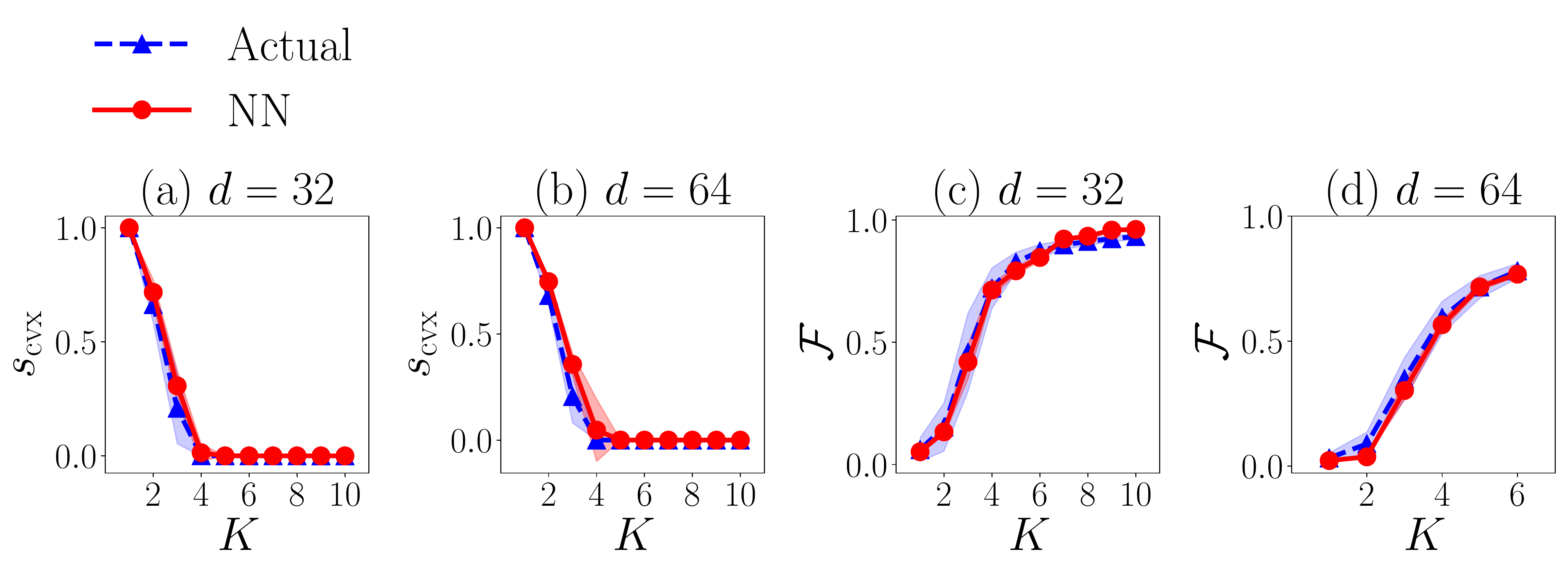}
	\caption{\label{fig:scvx_fid_d32_d64}Performances of ICCNet and FidNet for pure quantum systems of dimensions (a,c)~$d=32$ and (b,d)~$d=64$ obtained for the sake of generating Fig.~8 in the main text. Statistical noise from $N=1000$ copies per basis has been considered (refer to Tab.~\ref{tab:data}). In this case, the FidNet is trained to recognize fidelities with the ``right'' target states for simplicity, which is valid as no systematic errors are present here. FidNet training stops at $K=6$ for $d=64$ due to limited GPU resources.}
\end{figure*}

\begin{figure*}[htp]
	\centering\includegraphics[width=0.7\columnwidth]{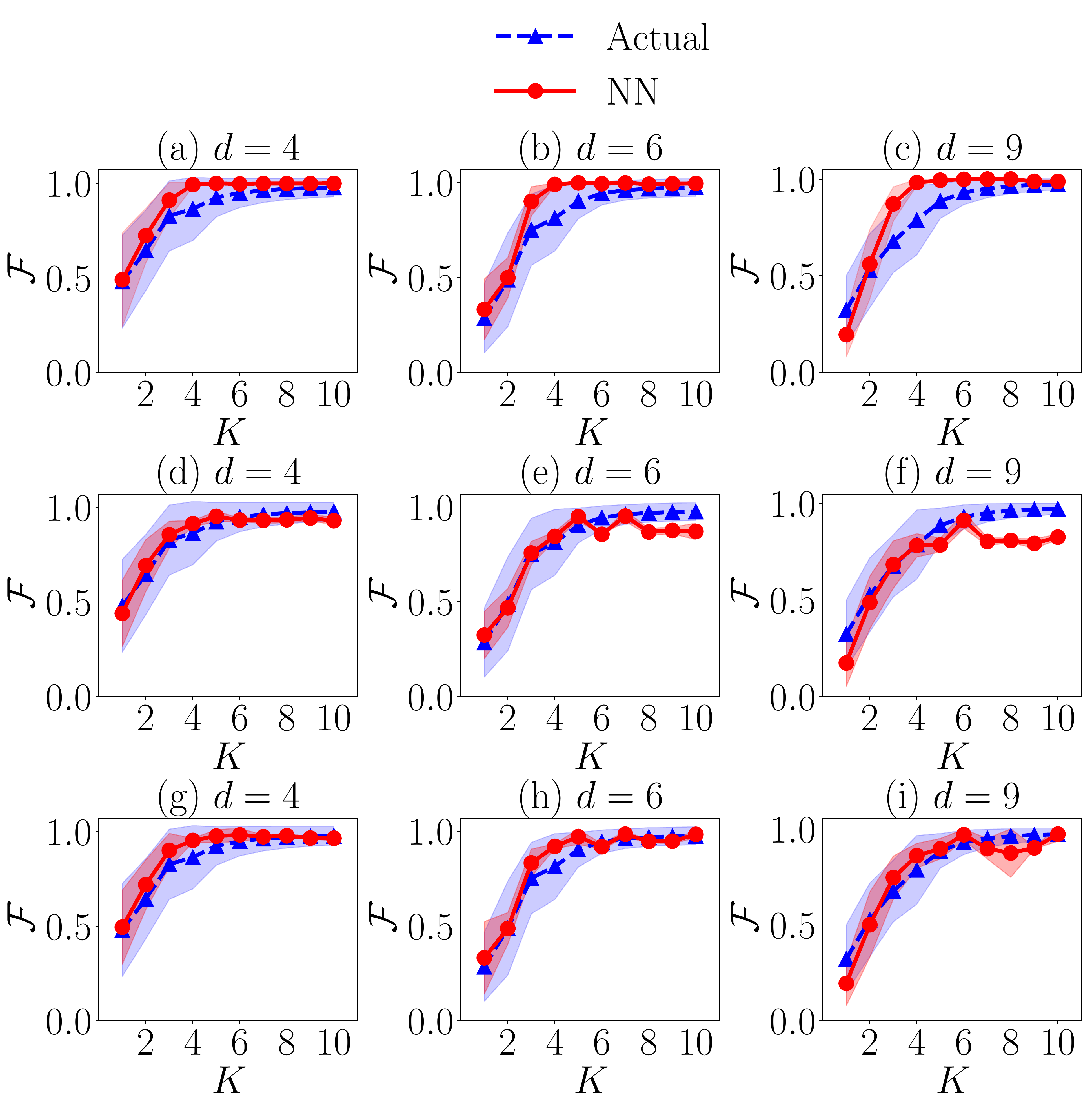}
	\caption{\label{fig:FidNet_compare}Performances of FidNet for benchmarking spatial-mode photonic datasets (a,b,c)~with the ``right'' target states, (d,e,f)~the ``wrong'' target states, and (g,h,i)~both types of target states. Average-fidelity benchmarking accuracies are higher when FidNet is trained with both ``right'' and ``wrong'' target states.}
\end{figure*}

\begin{figure*}[htp]
	\centering\includegraphics[width=0.7\columnwidth]{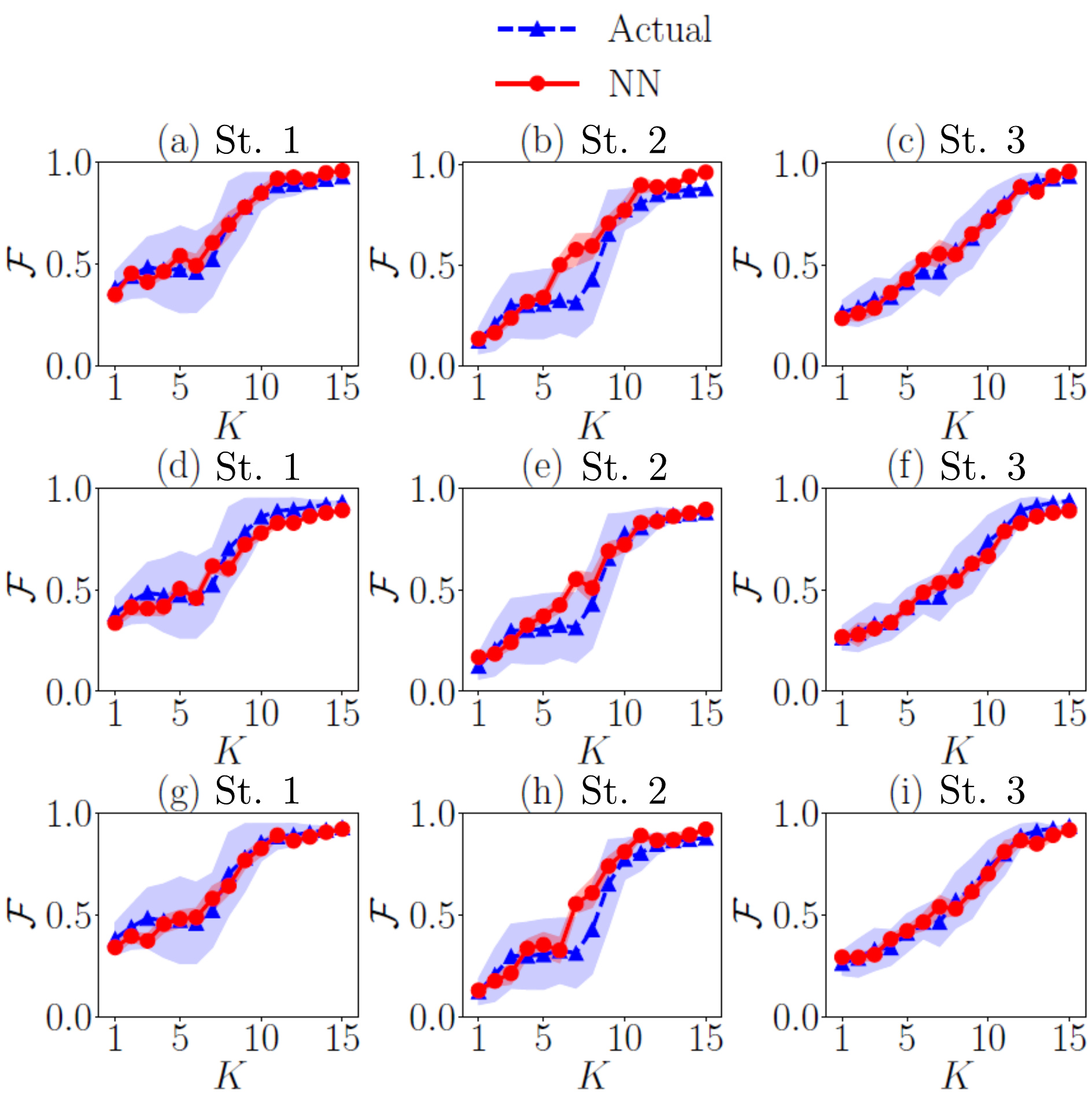}
	\caption{\label{fig:FidNet_compare2}Performances of FidNet for benchmarking three-photon datasets, where all specifications follow those of Fig.~\ref{fig:FidNet_compare}. For these three tested states, training with all the different types of target states give comparable accuracies.}
\end{figure*}

\begin{figure*}[h!]
	\centering\includegraphics[width=0.9\columnwidth]{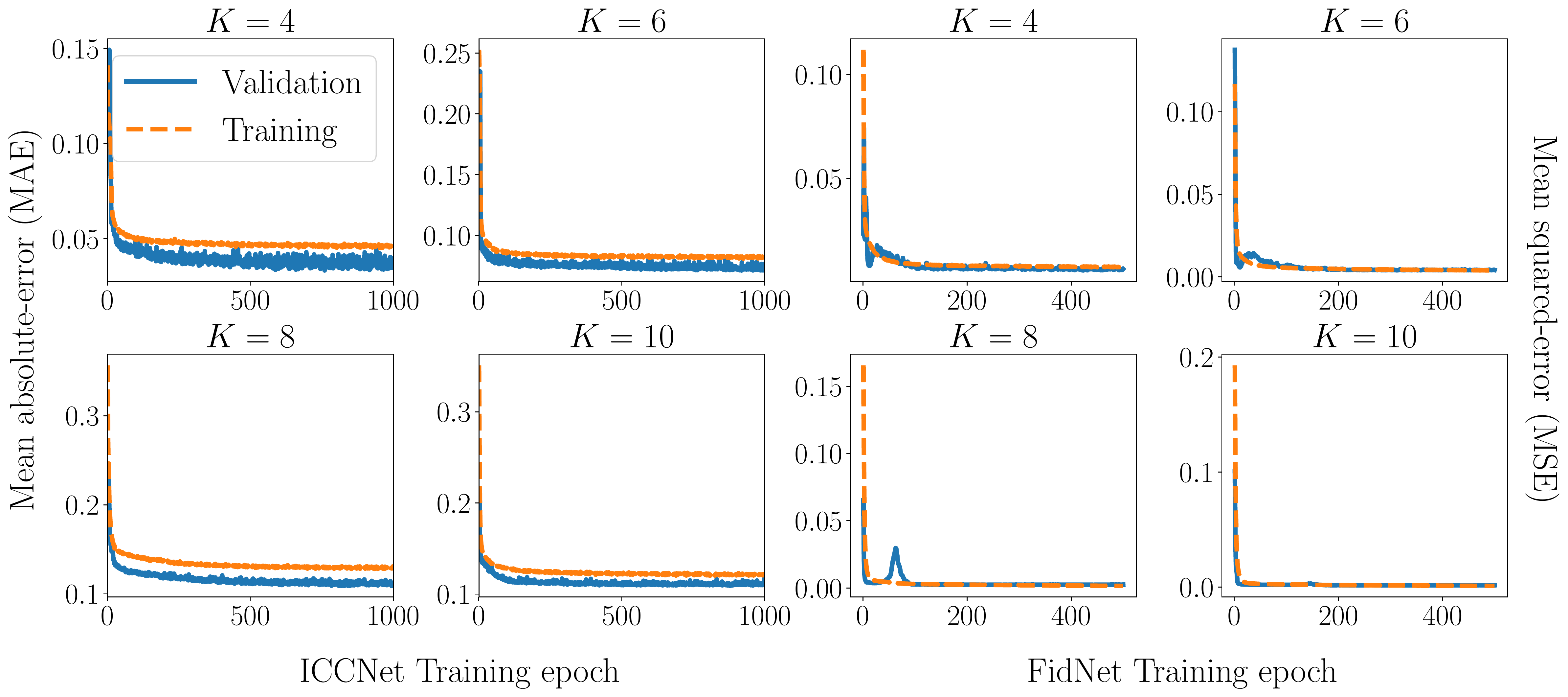}
	\caption{\label{fig:train_val}The progress of both ICCNet and FidNet training and validation loss values with the number of training epochs~(iterative steps) for $d=16$. All four data types (see Tab.~\ref{tab:data}) are stacked and trained using a common architecture for each Net. The loss values are computed with outputs $\rvec{y}_\text{train}$ and $\rvec{y}_\text{val}$. The trained models corresponding to the lowest validation loss are saved.}
\end{figure*}

\begin{figure*}[htp]
	\centering\includegraphics[width=0.7\columnwidth]{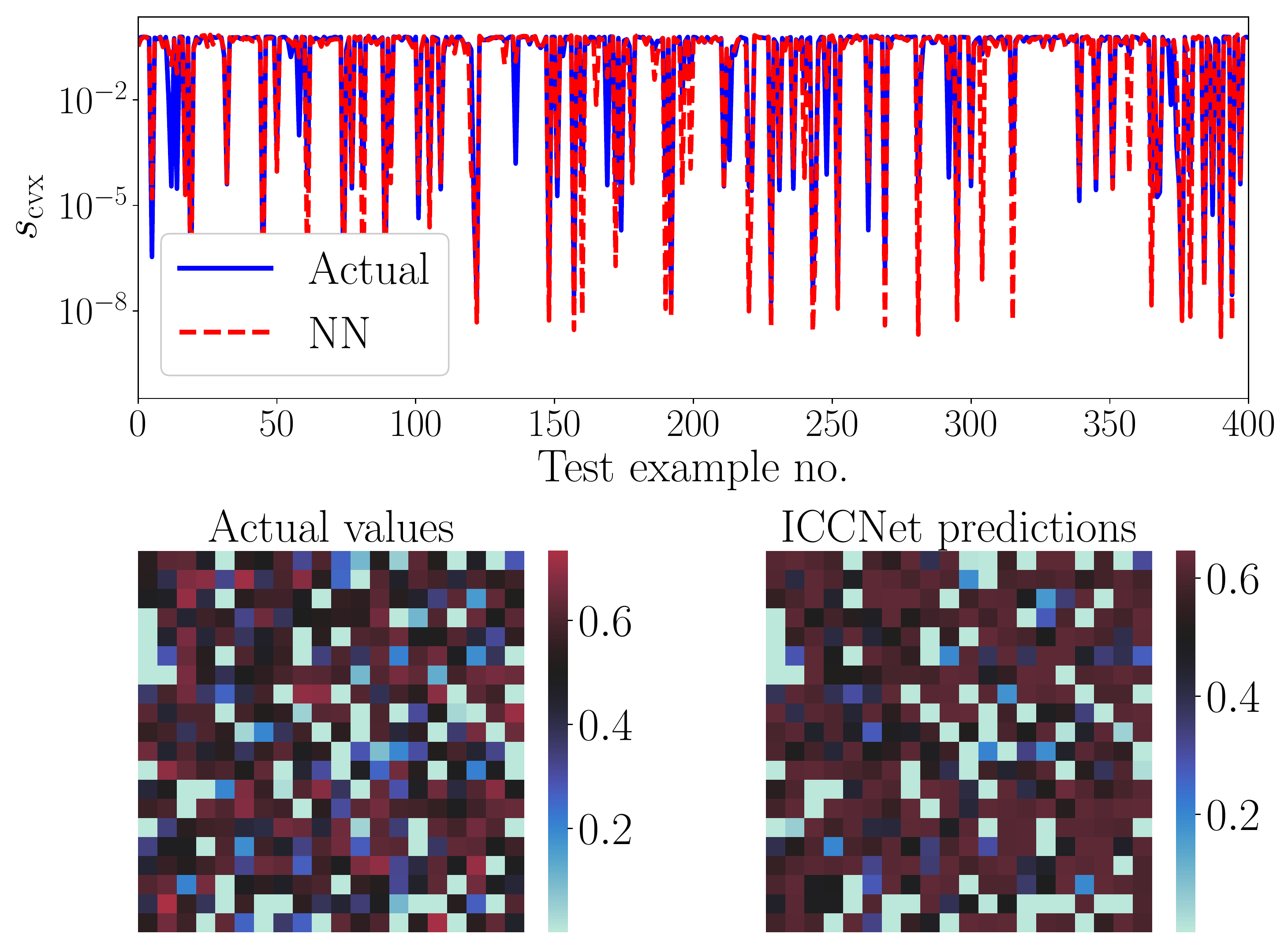}
	\caption{\label{fig:scvx_test}A sample of 400 $s_\textsc{cvx}$ test values of assorted datasets (a mixture of all the $d=16$ data types listed in Tab.~\ref{tab:data}) unseen during ICCNet training for $K=4$. The $20\times20$ heat maps that respectively represent these 400 actual and their corresponding ICCNet predicted~(NN) values serve to facilitate a more convenient visual comparison.} 
\end{figure*}

\begin{figure*}[htp]
	\centering\includegraphics[width=0.7\columnwidth]{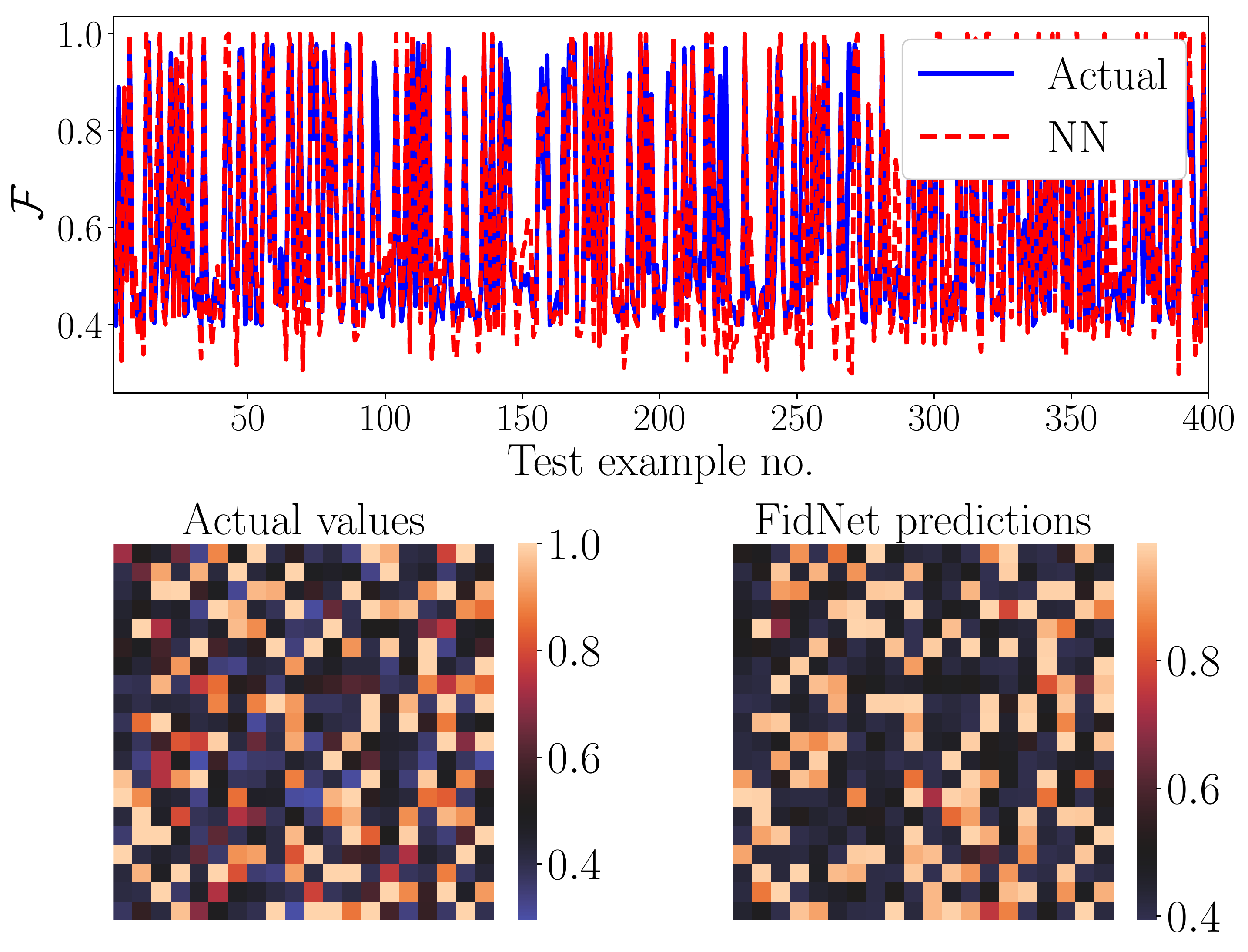}
	\caption{\label{fig:fid_test}A sample of 400 fidelity~($\mathcal{F}$) test values of assorted datasets (a mixture of all the $d=16$ data types listed in Tab.~\ref{tab:data}) unseen during FidNet training for $K=4$. The $20\times20$ heat maps that respectively represent these 400 actual and their corresponding FidNet predicted~(NN) values serve to facilitate a more convenient visual comparison.}
\end{figure*}

\begin{figure*}[htp]
	\centering\includegraphics[width=0.7\columnwidth]{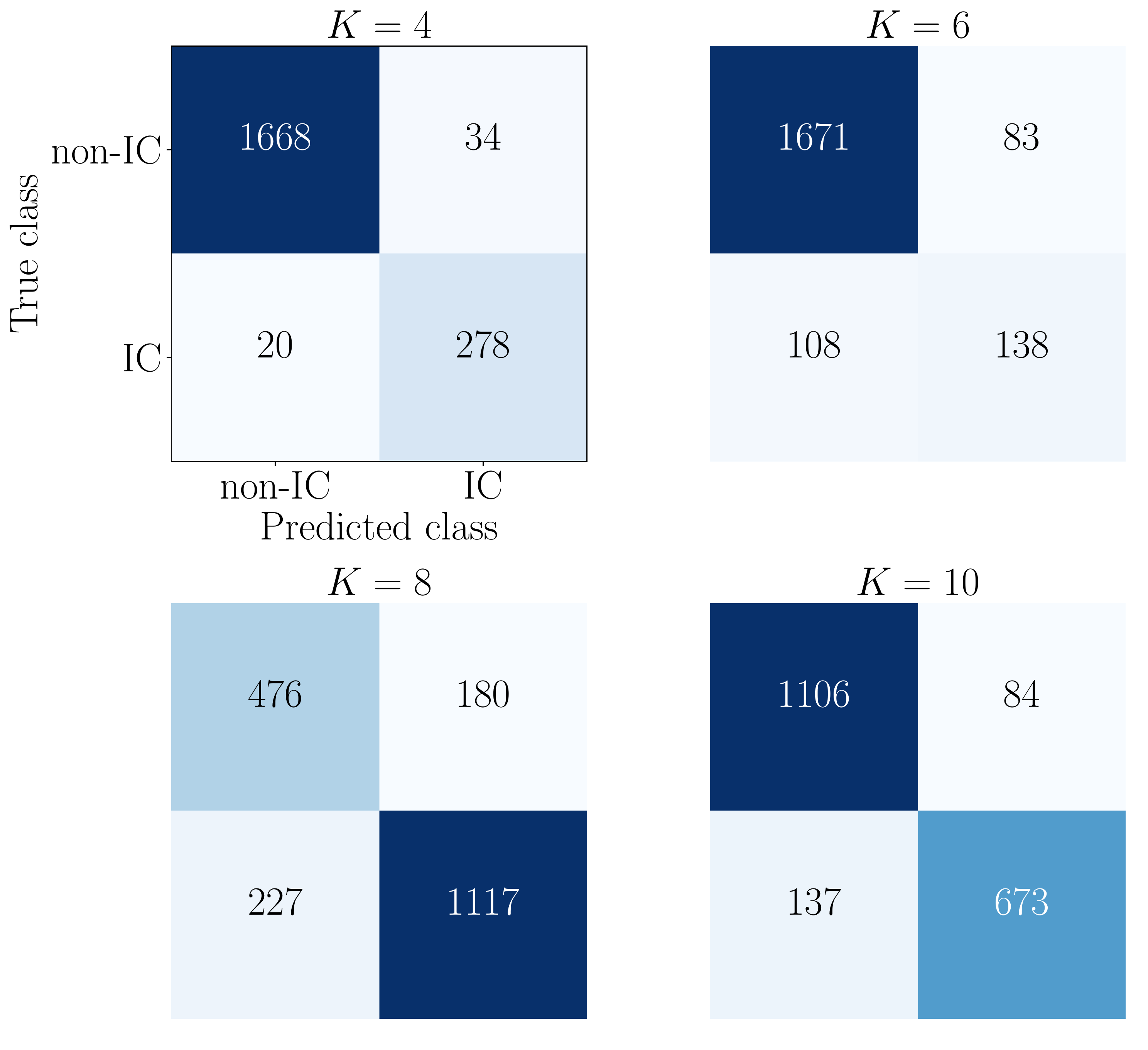}
	\caption{\label{fig:CM}The confusion matrix for $d=16$ and various $K$ values that classifies a total of 2000 test examples previously unseen by ICCNet. The larger the diagonals, the better the ICCNet prediction quality, as they represent successful predictions of correct classes of values. The respective F1~scores, in ascending order of $K$ shown here, are 0.911, 0.591, 0.846 and 0.859.}
\end{figure*}

\begin{figure*}[h!]
	\centering\includegraphics[width=1\columnwidth]{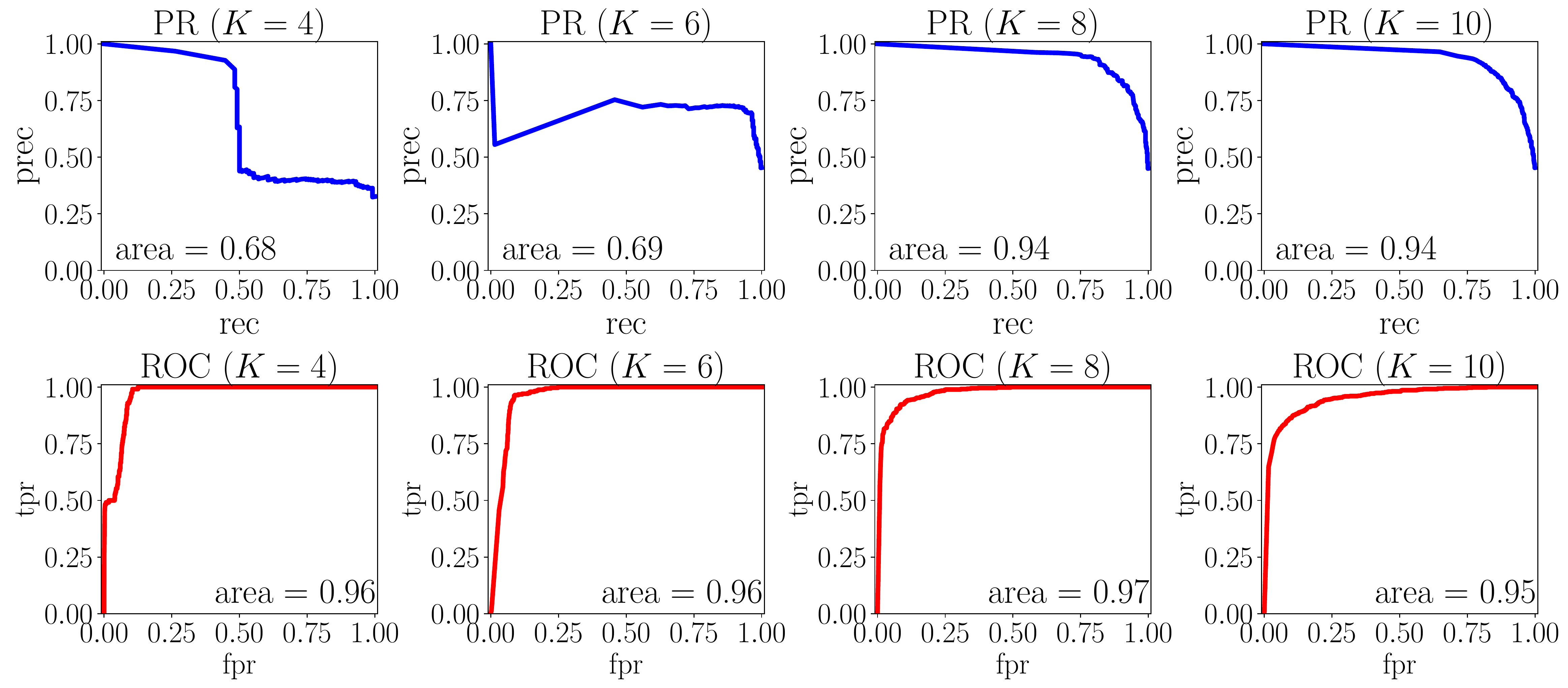}
	\caption{\label{fig:PR-ROC}Area under the PR and ROC curves.}
\end{figure*}

\end{document}